%
\magnification=\magstep1 \overfullrule=0pt
\advance\hoffset by -0.3truecm  
\font\tenmsb=msbm10       \font\sevenmsb=msbm7
\font\fivemsb=msbm5       \newfam\msbfam
\textfont\msbfam=\tenmsb  \scriptfont\msbfam=\sevenmsb
\scriptscriptfont\msbfam=\fivemsb
\def\Bbb#1{{\fam\msbfam\relax#1}}
\def\Z{{\Bbb Z}}

\font\grosss=cmr8 scaled \magstep4


\font\klein=cmr8 \font\itk=cmti8  \font\bfk=cmbx8   

\font\fat=cmmib10  
\font\smmfat=cmmib8 
\font\smfat=cmmib7 
\font\sf=cmss10   
%
\def\frac#1#2{{#1\over#2}}

%
\def\scst{\scriptstyle} \def\ssstyle{\scriptscriptstyle}
\def\test{\textstyle}   
%
\def\bn{\bigskip\noindent}   \def\mn{\medskip\smallskip\noindent}
\def\sn{\smallskip\noindent} \def\hbn{\hfill\break\noindent}
%

%
   
\def\cW{{\cal W}} \def\cH{{\cal H}} 
 
\def\one{{\bf 1}}

\def\fata{\hbox{{\fat\char"0B}}}  \def\sfata{\hbox{{\smfat\char"0B}}}
 
\def\fatb{\hbox{{\fat\char"0C}}} 
\def\fatl{\hbox{{\fat\char"15}}}
\def\fatm{\hbox{{\fat\char"16}}}

\def\sfatb{\hbox{{\smfat\char"0C}}} 
\def\sfatl{\hbox{{\smfat\char"15}}}
\def\sfatm{\hbox{{\smfat\char"16}}}

\def\bz{{\bar z}}   
 
 %

%
\def\lb{\lbrack}\def\rb{\rbrack}  
\def\q#1{\hbox{$\lb$\hskip.5pt{\rm #1}\hskip.5pt$\rb$}}

\def\tiNS{{\ssstyle\widetilde{\raise4pt\hbox{\fiverm NS}}}}
\def\tiR{{\ssstyle\widetilde{\raise4pt\hbox{\fiverm R}}}}
\def\lla{\langle\!\langle} \def\rra{\rangle\!\rangle}

\def\grapp{\hbox{\raise2.5pt\hbox{${\scst >}$}
\hskip-9pt\raise-2pt\hbox{${\scst \approx}$}}}
\def\uplc{^{L_0-{c\over 24}}}
\def\tqlc{{\tilde q\uplc}}
\def\oh{{1\over2}}

\def\sqr#1#2{{\vcenter{\vbox{\hrule height.#2pt
       \hbox{\vrule width.#2pt height #1pt \kern#1pt
         \vrule width.#2pt}\hrule  height.#2pt}}}}

\def\loarrr{\raise3.9pt\hbox{$\underline{\phantom{xxxxxx}}$}
\!\!\!\longrightarrow}
\def\loarrl{\longleftarrow\!\!\!\raise3.9pt\hbox{$
\underline{\phantom{xxxxxx}}$}}
\def\Fus#1#2#3#4#5#6{
 {\hbox{\sf F}}_{#1#2} \raise-.5pt\hbox{$\bigl\lb$}
 \raise.5pt\hbox{ ${\scst{ {\hfil\!#3\hfil\;\hfil#4\hfil} \atop
           {\hfil\!#5\hfil\;\hfil#6\hfil} }} $}
  \raise-.5pt\hbox{$\bigr\rb$}   }
\def\Fusmone#1#2#3#4#5#6{
 {\hbox{\sf F}}^{-1}_{#1#2} \raise-.5pt\hbox{$\bigl\lb$}
 \raise.5pt\hbox{ ${\scst{ {\hfil\!#3\hfil\;\hfil#4\hfil} \atop
           {\hfil\!#5\hfil\;\hfil#6\hfil} }} $}
  \raise-.5pt\hbox{$\bigr\rb$}   }
\def\Picfus#1#2#3#4#5#6{
 {\hbox{\sf F}}_{#1#2} \raise-.5pt\hbox{$\Bigl\lb$}
 \raise.5pt\hbox{ ${\test{ {\hfil\!#3\hfil\;\hfil#4\hfil} \atop
           {\hfil\!#5\hfil\;\hfil#6\hfil} }} $}
  \raise-.5pt\hbox{$\Bigr\rb$}   }\def\Brai#1#2#3#4#5#6{
 {\hbox{\sf B}}_{#1#2} \raise-.5pt\hbox{$\bigl\lb$}
 \raise.5pt\hbox{ ${\scst{ {\hfil\!#3\hfil\;\hfil#4\hfil} \atop
           {\hfil\!#5\hfil\;\hfil#6\hfil} }} $}
  \raise-.5pt\hbox{$\bigr\rb$}   }
\def\Picbrai#1#2#3#4#5#6{
 {\hbox{\sf B}}_{#1#2} \raise-.5pt\hbox{$\Bigl\lb$}
 \raise.5pt\hbox{ ${\test{ {\hfil\!#3\hfil\;\hfil#4\hfil} \atop
           {\hfil\!#5\hfil\;\hfil#6\hfil} }} $}
  \raise-.5pt\hbox{$\Bigr\rb$}   }
\def\cedille#1{\setbox0=\hbox{#1}\ifdim\ht0=1ex \accent'30 #1%
 \else{\ooalign{\hidewidth\char'30\hidewidth\crcr\unbox0}}\fi}
  
%

%
\global\newcount\glgnum \global\glgnum=0
\def\glg{{{\global\advance\glgnum by1}{(\number\glgnum)}}}
\def\mkg#1{\glg\xdef#1{(\the\glgnum)}}
\global\newcount\refnum \global\refnum=0
\def\preref{{\global\advance\refnum by1}{\q{\number\refnum}}}
\def\ref#1{\preref\xdef#1{{\the\refnum}}}
\def\eq{\eqno\mkg}
\global\newcount\kapnum
\global\newcount\glgnum
\global\newcount\thmnum
\global\newcount\chapflag
\global\newcount\footflag
\global\kapnum=0
\global\glgnum=0
\global\thmnum=0
\global\chapflag=0
\global\footflag=0
\xdef\cph{}
\xdef\kapsym{}
\def\chaphead#1{\xdef\cph{#1}}
\def\chapsym#1{\xdef\kapsym{#1}}
\def\chapter#1#2{{\chapsym{#1}\chaphead{#2}\bf 
\leftline{#2}}
               \global\advance\kapnum by1\global\chapflag=1
               \global\glgnum=0\global\thmnum=0\vskip-8pt\line{\hrulefill}\mn}
\def\thm{{{\global\advance\thmnum by1}{{\kapsym}.\number\thmnum}}}
\def\mkthm#1{\thm\xdef#1{{\kapsym}.\the\thmnum}}
%
\def\mkgho#1{\xdef#1{({\rm\kapsym}\the\glgnum}}
\def\mkgvo#1{\xdef#1{\the\glgnum)}}
\def\mkghvo#1{\xdef#1{\the\glgnum}}
\newread\epsffilein    \newif\ifepsffileok    \newif\ifepsfbbfound
\newif\ifepsfverbose   \newdimen\epsfxsize    \newdimen\epsfysize
\newdimen\epsftsize    \newdimen\epsfrsize    \newdimen\epsftmp
\newdimen\pspoints  \pspoints=1bp  \epsfxsize=0pt  \epsfysize=0pt
\def\epsfbox#1{\global\def\epsfllx{72}\global\def\epsflly{72}%
 \global\def\epsfurx{540}\global\def\epsfury{720}%
 \def\lbracket{[}\def\testit{#1}\ifx\testit\lbracket
 \let\next=\epsfgetlitbb\else\let\next=\epsfnormal\fi\next{#1}}%
\def\epsfgetlitbb#1#2 #3 #4 #5]#6{\epsfgrab #2 #3 #4 #5 .\\%
 \epsfsetgraph{#6}}%
\def\epsfnormal#1{\epsfgetbb{#1}\epsfsetgraph{#1}}%
\def\epsfgetbb#1{\openin\epsffilein=#1
\ifeof\epsffilein\errmessage{I couldn't open #1, will ignore it}\else
 {\epsffileoktrue \chardef\other=12
  \def\do##1{\catcode`##1=\other}\dospecials \catcode`\ =10 \loop
  \read\epsffilein to \epsffileline \ifeof\epsffilein\epsffileokfalse\else
       \expandafter\epsfaux\epsffileline:. \\    \fi
  \ifepsffileok\repeat   \ifepsfbbfound\else
  \ifepsfverbose\message{No bounding box comment in #1; using defaults}\fi\fi
  }\closein\epsffilein\fi}%
\def\epsfclipstring{}%

\def\epsfsetgraph#1{ \epsfrsize=\epsfury\pspoints
 \advance\epsfrsize by-\epsflly\pspoints  \epsftsize=\epsfurx\pspoints
 \advance\epsftsize by-\epsfllx\pspoints \epsfxsize\epsfsize\epsftsize
  \epsfrsize
\ifnum\epsfxsize=0\ifnum\epsfysize=0\epsfxsize=\epsftsize\epsfysize=\epsfrsize
   \epsfrsize=0pt \else\epsftmp=\epsftsize \divide\epsftmp\epsfrsize
  \epsfxsize=\epsfysize \multiply\epsfxsize\epsftmp
  \multiply\epsftmp\epsfrsize \advance\epsftsize-\epsftmp \epsftmp=\epsfysize
  \loop \advance\epsftsize\epsftsize \divide\epsftmp 2 \ifnum\epsftmp>0
  \ifnum\epsftsize<\epsfrsize\else \advance\epsftsize-\epsfrsize
  \advance\epsfxsize\epsftmp \fi \repeat \epsfrsize=0pt \fi \else
  \ifnum\epsfysize=0 \epsftmp=\epsfrsize \divide\epsftmp\epsftsize
  \epsfysize=\epsfxsize \multiply\epsfysize\epsftmp
  \multiply\epsftmp\epsftsize \advance\epsfrsize-\epsftmp \epsftmp=\epsfxsize
  \loop \advance\epsfrsize\epsfrsize \divide\epsftmp 2 \ifnum\epsftmp>0
  \ifnum\epsfrsize<\epsftsize\else \advance\epsfrsize-\epsftsize
  \advance\epsfysize\epsftmp \fi \repeat \epsfrsize=0pt \else
  \epsfrsize=\epsfysize \fi \fi \ifepsfverbose\message{#1:
  width=\the\epsfxsize, height=\the\epsfysize}\fi \epsftmp=10\epsfxsize
  \divide\epsftmp\pspoints \vbox to\epsfysize{\vfil\hbox to\epsfxsize{
  \ifnum\epsfrsize=0\relax \includegraphics{#1} \else \epsfrsize=10\epsfysize \divide\epsfrsize\pspoints
  \includegraphics{#1}\fi \hfil}}%
\global\epsfxsize=0pt\global\epsfysize=0pt}%
{\catcode`\%=12 \global\let\epsfpercent=
\long\def\epsfaux#1#2:#3\\{\ifx#1\epsfpercent
   \def\testit{#2}\ifx\testit\epsfbblit    \epsfgrab #3 . . . \\%
      \epsffileokfalse     \global\epsfbbfoundtrue
   \fi\else\ifx#1\par\else\epsffileokfalse\fi\fi}%
\def\epsfempty{}\def\epsfgrab #1 #2 #3 #4 #5\\{%
\global\def\epsfllx{#1}\ifx\epsfllx\epsfempty \epsfgrab #2 #3 #4 #5 .\\\else
   \global\def\epsflly{#2} \global\def\epsfurx{#3}\global\def\epsfury{#4}\fi}
\def\epsfsize#1#2{\epsfxsize} 
\setbox82=\vbox{\phantom{
\ref{\Caone}\ref{\Ca}\ref{\Lew}\ref{\CaL}\ref{\PSS}
\ref{\Run}\ref{\BPPZ}\ref{\FFFS}
\ref{\JoseSonia}
\ref{\Costasetal}
\ref{\FS}
\ref{\RSm}
\ref{\KS}\ref{\BHS}
\ref{\MattTerr}\ref{\Ban}
\ref{\FM}\ref{\ncg}\ref{\Antonetal}
\ref{\Gep}\ref{\FKS}\ref{\OOY}\ref{\RSg}\ref{\BDLR}\ref{\CY}
\ref{\DouCat}\ref{\BS}\ref{\FSW}\ref{\BDi}\ref{\DFi}
\ref{\MattSaku}\ref{\PZ}\ref{\Scheid}
\ref{\GJS}\ref{\HIV}\ref{\Kenn}
}}
{\nopagenumbers
\line{hep--th/0208119 \hfill KCL-MTH-01-45, ESI 1193 (2002)}
\vskip3cm
\centerline{\grosss Permutation Branes}
\vskip1.3cm
\centerline{\bf Andreas Recknagel}
\vskip.7cm
\centerline{Department of Mathematics, King's College London}
\smallskip
\centerline{Strand, London WC2R 2LS, UK}
\smallskip
\centerline{\tt anderl@mth.kcl.ac.uk}
\vskip1.8cm
\centerline{\bf Abstract}
\bigskip

{\narrower\narrower \noindent
$N$-fold tensor products of a rational CFT carry an action of the
permutation group $S_N$. These automorphisms can be used as gluing 
conditions in the study of boundary conditions for tensor product 
theories. We present an ansatz for such permutation boundary 
states and check that it satisfies the cluster condition and Cardy's 
constraints. For a particularly simple case, we also investigate 
associativity of the boundary OPE, and find an intriguing connection 
with the bulk OPE. In the second part of the paper, the constructions 
are slightly extended for application to Gepner models. We give 
permutation branes for the quintic, together with some formulae 
for their intersections. 

} 
\bigskip 
  
\vfil
\centerline{\it In memory of}
\smallskip
\centerline{\it Sonia Stanciu}
\medskip
\phantom{xxx}
\eject}
\pageno=1
\noindent
{\bf 1. Introductory remarks}
\mn
Thanks to the pioneering work by Cardy \q{\Caone,\Ca} and to further, 
more recent developments (see e.g.\ \q{\Lew-\FFFS}), we have a 
reasonably complete understanding of a certain class of boundary 
conditions for rational conformal field theories. This class is 
distinguished by an especially simple relation between the form 
of the bulk partition function and the gluing automorphism 
$\Omega$ that connects left- and right-moving symmetry generators
$$ W(z) = \Omega\, \overline{W}{}(\bz)
\quad\ {\rm for}\ \ z = \bar z \ \; :
$$
If the rational bulk theory comes with a charge-conjugate modular 
invariant partition function 
$$
Z(q,\bar q) = \sum_{i \in {\cal I}} \;
          \chi_{i}(q)\,\chi_{i^+}(q)^* \ \;, 
$$
it is the standard gluing condition $\Omega = {\rm id}$ that leads 
to a simple picture of boundary conditions, basically because all 
Ishibashi states that can be formed for standard gluing conditions 
are actually present in the bulk theory. The same is true for 
general $\Omega$ as long as, in the bulk partition function $Z(q,\bar q)$, 
each sector $i$ is paired with $\omega^{-1}(i^+)$ instead of the 
charge-conjugate sector $i^+$, where $\omega(i)$ labels the 
representation of the symmetry algebra $\cW$ induced in the sector $i$
by the action of $\Omega$. The boundary states associated with this 
class of boundary conditions are often referred to as ``Cardy 
boundary states''.  
\hbn
Non-standard gluing conditions (for bulk theories whose partition 
function is not of the appropriate $\omega$-type) are of crucial 
importance in string theory, starting with Dirichlet boundary conditions 
for free bosons. The non-trivial automorphisms $\Omega$ available for 
the gluing conditions of a rational CFT of course depend on the symmetry 
algebra $\cW$ of the model. But there is  a simple and natural 
construction that yields, from a given rational model, new ones 
on which universal automorphism groups act -- namely taking tensor 
products of the original ``component theory''. While this operation 
alone does not produce very exciting effects in the domain of bulk 
theories, the structure of associated boundary CFTs is much richer: 
The set of all boundary conditions for an $N$-fold tensor product, 
even those that preserve the full symmetry, is not simply obtained by 
forming tensor products of boundary conditions for the component 
theory, but includes boundary conditions associated with 
gluing automorphisms from the permutation group $S_N$. 
\sn
In this paper, we present general formulae for such ``permutation 
boundary states'', first focusing on situations where the original
RCFT has a charge-conjugate (or, if all sectors are self-conjugate, a
diagonal) partition function. The permutation boundary states are 
still ``rational'' in the sense that the boundary CFT is covariant 
under the full symmetry algebra $\cW^N$, but we will see that 
the excitation (or open string) spectra of permutation branes in 
general look rather different from those of tensor products of 
Cardy states. 
Our work generalises studies of permutation branes in $G \times G$ 
WZW models \q{\JoseSonia}. On the other hand, our boundary states 
are special cases of the ``conformal walls'' that appeared in 
\q{\Costasetal}:\ while a general conformal wall is nothing but 
a symmetry-breaking conformal boundary condition for a tensor 
product model (not necessarily with identical factors), our permutation 
branes can be viewed as walls with perfect transmission 
(or reflection) of energy between (or within) the identical component 
theories. 
\mn
In the next section, we first specify some notations and convenient 
assumptions, analyse the set of permutation Ishibashi states and 
give an ansatz for the permutation boundary states. Then, two types
of non-linear constraints are checked for this ansatz, namely the
cluster condition in Subsection 2.2 and Cardy's constraints in 
Subsection 2.3. The latter contains explicit results for spectra 
of open strings supported by one permutation brane or stretching 
between two different ones. Up to now, a third important set of 
sewing relations, namely associativity of the operator product 
expansion of boundary fields, can only be discussed in a rather 
simple case (a specific brane for a two-fold tensor product) 
-- which however features an unexpected connection between the 
boundary OPE on this permutation brane and the OPE of the 
component bulk CFT. 
\hbn
The third section is devoted to the construction of permutation branes 
for a CFT of interest to string theory, namely the Gepner model 
corresponding to the quintic Calabi-Yau manifold. Gepner models are 
constructed from tensor products of $N=2$ superconformal minimal 
models, and many of them admit an action of a permutation group
($S_5$ for the quintic). Because of various orbifold-like projections 
involved in the Gepner construction, the bulk partition functions of 
these models do not satisfy the assumptions used in Section 2, so 
some of the methods presented there have to be adapted. The Gepner 
projections are such that the Ishibashi state content depends 
strongly on the details of the models (the relative prime factors 
in minimal model levels and cycle lengths of the permutation), 
therefore we restrict ourselves to the quintic when computing 
explicit expressions for partition functions and intersection 
forms. From the former, one can read off the spectrum of 
massless open string states,  and they indicate that the new 
branes are BPS as expected. The intersection forms should 
set the stage for a geometric interpretation of permutation boundary 
states. This is, however, left as one of several open problems, which 
are listed in the concluding section. 

\bn\mn
{\bf 2. Permutation branes for rational CFTs with diagonal bulk invariant}
\mn
We start from a unitary rational CFT on the plane with chiral (left- and 
right-moving) symmetry algebras $\cW_L\ \simeq\ \cW_R\ \simeq\ \cW$, and 
with a charge-conjugate partition function 
$Z(q,\bar q) = \sum_{i \in {\cal I}} \chi_{{}_i}(q)\, \chi_{{}_{i^+}}(q)^*$ 
associated with a decomposition $\cH = \bigoplus_{i \in {\cal I}}\cH_i 
\otimes \cH_{i^+}$  of the state space into irreducibles. For notational
convenience, let us assume that all sectors are self-conjugate, i.e.\ 
that the fusion rules satisfy $N_{ii}^0 =1$ where 0 denotes the vacuum 
sector. Furthermore, it will at a certain point be advantageous to 
assume that no fusion 
channel with non-trivial multiplicity occurs:\ $N^k_{ij}=0$ or 1. 
\hbn
We can then form an $N$-fold tensor product of this rational CFT, with 
chiral algebra 
$\cW^N := \cW \times \cdots \times \cW$ and partition function 
$$
Z^{(N)}(q,\bar q) = \bigl(\,Z(q,\bar q)\,\bigr)^N 
           = \sum_{I \in {\cal I}^N} \chi_{{}_I}(q)\,\chi_{{}_I}(q)^*
$$ 
where we have used the multi-index $I := (i_1, \ldots, i_N)$ to label the 
characters of the tensor product theory, 
$\chi_{{}_I}(q) :=  \chi_{{}_{i_1}}\!(q)\cdots \chi_{{}_{i_N}}\!(q)$. 
Again, the partition function $Z^{(N)}(q,\bar q)$ is a diagonal 
(and at the same time, because of our simplifying assumption $i = i^+$, 
 a charge conjugate) modular invariant. 
\sn
The tensor product theory admits outer automorphisms which 
act by permutations of the components of $\cW^N$, namely 
$$
\Omega_\pi \;:\ W^{[k]}(z) \longmapsto W^{[\pi(k)]}(z)
$$ 
where $\pi \in S_N$ is a permutation and where $ W^{[k]}(z) := 
\one \otimes \cdots \otimes W(z) \otimes  \cdots \otimes \one$ 
denotes the action of a $\cW$-generator in the $k^{\rm th}$ component 
theory, for $k=1,\ldots,N$. We want to construct boundary conditions 
for the tensor product theory with $\Omega_\pi$ appearing as a gluing 
automorphism, i.e.\ we link left- and right-moving generators by the 
condition 
$$
 W^{[k]}(z) = \overline{W}{}^{[\pi(k)]}(\bz) 
\eq\permglu$$
along the boundary $z = \bz$ of the upper half-plane. These 
boundary conditions are conformal since $\Omega_\pi$ leaves the 
diagonal energy-momentum tensor $T = T^{[1]} + \ldots + T^{[N]}$ 
fixed. The gluing conditions \permglu\ in fact guarantee that the 
full symmetry algebra $\cW^N$ is represented on the Hilbert space 
of the boundary CFT (with constant boundary condition along the 
real line) even though the analytic continuation from upper to 
lower half-plane prescribed by \permglu\ is not the standard one. 
\bn
{\bf 2.1 An ansatz for permutation boundary states}
\mn
Permutation boundary states are built up from objects that 
implement the permutation gluing conditions \permglu. These 
Ishibashi states can be expanded as 
$$
|I\rra_\pi = \sum_{{\hbox{\smfat M}}}\ 
   |i_1,M_1\rangle  \otimes \cdots \otimes |i_N,M_N\rangle 
   \otimes U|i_{\pi^{-1}(1)},M_{\pi^{-1}(1)}\rangle  
    \otimes \cdots \otimes U |i_{\pi^{-1}(N)},M_{\pi^{-1}(N)}\rangle 
\eq\permIshi$$
where {\fat M} $= (M_1,\ldots,M_N)$ is used to label orthonormal bases 
$|i_k,M_k\rangle$  of energy eigenstates in the representations $\cH_{i_k}$ 
of $\cW$, and where the operator $U$ in front of the right-movers 
is the chiral CPT operator as usual. It is important to realize 
that the objects $|I\rra_\pi$ are available only for certain 
multi-indices $I = (i_1, \ldots, i_N)$: Since the partition 
function of the bulk theory is diagonal, 
$$ 
|I\rra_\pi\quad {\rm exists\ if\ and\ 
only\ if}\quad 
i_k = i_{\pi^{-1}(k)}\quad {\rm for\ all}\ \, k = 1,\ldots,N\ \ .
\eq\Ishiexcond$$ 
This means that the two 
$\cW$-representations $i_k,\,i_l$ have to coincide whenever $k$ and 
$l$ are elements of the same cycle of the permutation  
\def\mone{\mathop{\raise1.5pt\hbox{\vbox{\hrule width 1.7pt height.4pt}}}}
\def\upi{^{\hskip-1.7pt{\scriptscriptstyle \pi}}}
\def\uupi{\hskip-0.8pt\raise5pt\hbox{${\scriptscriptstyle \pi}$}}
\def\usi{^{\hskip-1.7pt{\scriptscriptstyle \sigma}}}
\def\uusi{\hskip-1.1pt\raise5pt\hbox{${\scriptscriptstyle \sigma}$}\hskip-4pt}
\def\uupsi{\hskip-4pt\raise5pt\hbox{
          ${\scriptscriptstyle \pi^{\mone\hskip-.2pt1}\!\sigma}$}}
\def\upsi{^{\hskip-1.7pt{\scriptscriptstyle \pi^{\mone\hskip-.2pt1}\!\sigma}}}
\def\dowpsi{_{\hskip-1.7pt{\scriptscriptstyle \pi^{\mone\hskip-.2pt1}\!\sigma}}}
$$\eqalign{
\pi = \bigl(\,n\upi_1\ \pi(n\upi_1)\;&\ldots\;
                     \pi^{\Lambda\upi_1-1}(n\upi_1)\,\bigr)\  
\bigl(\,n\upi_2\ \pi(n\upi_2)\;\ldots\;\pi^{\Lambda\upi_2-1}(n\upi_2)\,\bigr)\ 
\cr
&\qquad\qquad\qquad \ldots\quad \bigl(\,n\upi_{\!P\upi}\ 
        \pi(n\upi_{\!P\upi})\;\ldots\;
        \pi^{\Lambda\upi_{\!P\upi}\,-1}(n\upi_{P\upi})\,\bigr)\ \ .
\cr}\eq\picycles$$
Here, we have chosen an arbitrary element $n\upi_\nu \in \{1,\ldots,N\}$   
as representative of the $\nu^{\rm th}$ cycle 
$$
 C\uupi_\nu = \bigl(\,n\upi_\nu\ \pi(n\upi_\nu)\;\ldots\;
                     \pi^{\Lambda\upi_\nu-1}(n\upi_\nu)\,\bigr) \ \ ;
$$
the $n\upi_\nu$ will be kept fixed, and we denote the length of  $C\uupi_\nu$ 
by $\Lambda\uupi_\nu$, for $\nu = 1,\ldots,P\uupi$, where $P\uupi$ is the 
number of cycles of $\pi$. 
\hbn
\vbox{\noindent
We will abbreviate the condition $i_k = i_{\pi^{-1}(k)}$ on the 
Ishibashi states by inserting a Kronecker symbol $\delta^{C^\pi}_I$ 
into the following formulae. Note that, even though $i_k=i_l$, the 
summation indices $M_k,\;M_l$ in \permIshi\  are independent of each 
other -- which in particular makes it clear that the permutation 
Ishibashi states are not just superpositions of standard ones. 
\sn
Full-fledged boundary states $|\!|\,{\fata}\,\rra_\pi$ for the gluing 
conditions \permglu\ can be written as linear combinations of the 
Ishibashi states \permIshi, 
$$
|\!|\,{\fata}\,\rra_\pi = \sum_I\ \delta^{C^\pi}_I\ B^I_{{\sfata}}\;
|I\rra_\pi\ \ .
\eq\permbdst$$
We will now present a natural ansatz for the coefficients $B^I_{{\sfata}}$ 
and then perform two important consistency checks, involving the 
so-called cluster condition resp.\ Cardy's conditions on the
strip partition functions. 
\hbn
In our ansatz for the coefficients in \permbdst, $\fata = (\alpha_1,\ldots, 
\alpha_{n_P})$ is a multi-index with as many components as there are 
independent labels $i_{n\upi_\nu}$ in the Ishibashi states $|I\rra_\pi$; each 
$\alpha_\nu$ is taken from the label set ${\cal I}$ of all 
$\cW$-representations. The formula for $B^I_{{\sfata}}$ is multiplicative 
in the  $i_{n\upi_\nu}$ and reduces to Cardy's 
solution for the component theory in the special case $N=1$: 
$$
B^I_{{\sfata}} = B_{\alpha_1}^{i_{n_1}} \cdots 
            B_{\alpha_P}^{i_{n_P}}  \quad\quad  {\rm with}\quad\ \  
B_{\alpha_\nu}^{i_{n_\nu}} =  {S_{\alpha_\nu\,i_{n_\nu}}\over\bigl( 
                S_{0\,i_{n_\nu}}\bigr)^{{\Lambda_\nu\over2}}}\ \ ;
\eq\ansatz$$
we have dropped the superscript ${}\upi$ from the cycle representatives 
$n\upi_\nu$, the cycle lengths $\Lambda\uupi_\nu$ and the number of 
cycles $P\uupi\,$. As usual, the matrix $S$ implements modular 
transformation of the $\cW$-characters; therefore, the matrix 
$\bigl(\,B^I_{{\sfata}}\bigr)$ is invertible, and the 
ansatz \ansatz\ will provide a complete set, in the sense of \q{\PSS}, 
of boundary states for fixed gluing condition $\Omega_\pi$ 
-- provided all sewing constraints are satisfied.  
\bn
{\bf 2.2  Cluster condition}
\mn
The cluster condition gives a first check for the consistency of our 
ansatz. In contrast to Cardy's constraints, it involves a single 
boundary condition and genus zero world-sheets only. It is obtained from 
a sewing relation which compares two different ways (orderings of OPEs) 
to evaluate a bulk field two-point function in the presence of the 
boundary (see e.g.\ \q{\Lew,\PSS,\FS,\Run,\RSm}), with subsequent 
projection on the identity channel -- which governs the long range 
behaviour (the clustering properties) of the two-point function.
(Strictly speaking, this requires the boundary condition to be 
`fundamental' in the sense that only a single vacuum character is 
present in the open string spectrum; we will see in the next subsection 
that our permutation branes have this property.) The cluster condition 
reads
$$
B^I_{{\sfata}} \,B^J_{{\sfata}} 
    = \sum_K \ \Xi_{IJK}\; B^0_{{\sfata}}\, B^K_{{\sfata}} \ \ ,
\eq\cluster$$
where $\Xi_{IJK}$ is a product $C\cdot${\sf F} involving a structure 
constant $C$ from the OPE of the two bulk fields $\varphi_{I,I}$ and 
$\varphi_{J,J}$, as well as a certain element {\sf F} of the fusing 
matrix (which relates the conformal blocks in the two channels). 
Since these data are known explicitly only for few CFTs, the 
$\Xi_{IJK}$  can usually not be determined directly. 
}
\vbox{\noindent
However, for the component theory ($N=1$), a simple expression for 
the $\Xi_{IJK}$ follows from the Verlinde formula and the fact that 
Cardy's boundary states provide a complete set of consistent boundary 
states for standard gluing conditions (see in particular \q{\BPPZ,\FFFS}). 
Plugging in Cardy's solutions for the $B_\alpha^i$ into \cluster\ 
for $N=1$ yields 
$$
\Xi_{ijk} = \Bigl( {S_{00}S_{0k}\over S_{0i}S_{0j}}\Bigr)^{\oh}\,
       N_{ij}^k\ \ .
\eq\clusterconst$$
Clearly, the constants $\Xi_{IJK}$ for tensor product bulk fields 
$\varphi_I = \varphi_{i_1} \otimes \cdots \otimes\varphi_{i_N}$ etc.\ 
factorise into $N$ constants $\Xi_{i_l j_l k_l}$. The constraint 
\cluster\ implicitly requires that all the bulk fields involved 
have non-vanishing one-point functions. In particular, the bulk OPE 
on the rhs is usually truncated by taking $\langle\,\cdot\,\rangle_\alpha$. 
Thus, in our situation, only those tensor products occur that obey the 
$\delta_I^{C\upi}$-restriction from the Ishibashi states, and we have 
$$
\Xi_{IJK} = \bigl(\Xi_{  i_{n_1} j_{n_1} k_{n_1}  } \bigr)^{\Lambda_1} 
   \cdots \;  
     \bigl(\Xi_{i_{n_{\!P\upi}}\, j_{n_{\!P\upi}}\, k_{n_{\!P\upi}} } 
                  \bigr)^{\Lambda_{P\upi}} \ \ .
$$
As a result, the summation in \cluster\ splits into $P\uupi\;$ 
independent ones (with each $k_{n_\nu\upi}$ ranging over the full 
index set ${\cal I}$). Our 
simplifying assumptions on the fusion rules imply that 
$(N_{ij}^k)^M = N_{ij}^k$ for all $M \neq 0$; combining this with the fact 
that the generalised quantum dimensions represent the fusion rules, 
$$
{S_{ia}\over S_{0a}}\; {S_{ja}\over S_{0a}} = 
\sum_k\ N_{ij}^k\ {S_{ka}\over S_{0a}}\ \ ,
\eq\qdimfus$$
it is an easy exercise to check that the boundary states \ansatz\ 
indeed satisfy the cluster condition. 
It appears to be mainly a problem of notation to include non-trivial 
multiplicities of fusion channels: As in  \q{\BPPZ,\FFFS}, one would 
have to introduce additional block labels to disentangle summations.
In the following, the simplifying property $N_{ij}^k<2$ will play 
no role. 
\bn
{\bf 2.3 Cardy's conditions}
\mn
The computations necessary to test Cardy's conditions \q{\Ca} provide 
much more interesting physical information than the cluster condition. 
Cardy's conditions involve pairs of boundary states and require that 
the quantity 
$$
Z_{\sfata\sfatb} (q) :=\lla\, \fatb\, |\!|\, \tqlc\, |\!|\,{\fata}\,\rra
$$
can be interpreted as the partition function of a CFT on the strip
with boundary conditions $\fata$ resp.\ $\fatb$ along the two
boundaries; this partition function records the spectrum of the 
boundary CFT, or of excitations of open strings attached to the 
branes $\fata$ and $\fatb$. The actual condition hidden in this 
statement is that $Z_{\sfata\sfatb} (q) = \sum_k\, n_{\sfata\sfatb}^k\, 
\chi_{{}_k}^\circ(q)\,$ must be a linear combinations of characters 
$\chi_{{}_k}^\circ(q)$ for some conformal symmetry algebra 
$\cW^\circ$ with (positive) {\sl integer} coefficients 
$n_{\sfata\sfatb}^k$. We use $\tilde q = \exp(-2\pi i/\tau)$ 
and $q=\exp(2\pi i \tau)$, with $\tau$ being the modular parameter. 
}
\noindent
In order to check Cardy's constraints, we first
need to compute ``sandwiches'' of the closed string propagator between 
two Ishibashi states, possibly corresponding to two different 
permutations $\pi,\sigma\in S_N$: 
$$\eqalign{
{}_\sigma\lla I'|\, \tilde q^{L_0^{\rm tot} - {c^{\rm tot}\over24}}\,
                |I\rra_\pi 
&= \sum_{\hbox{\smfat M},\hbox{\smfat M}'}\ \;
\tilde q^{h(i_1,M_1)+\ldots+ h(i_N,M_N) - N {c\over24}}  
\cr
 \noalign{\vskip-10pt}  
&\hphantom{xxxxxxxxxxx} \times\ 
\langle i_1',M_1'\,|\,i_1,M_1\rangle\  \cdots \ 
       \langle i_N',M_N'\,|\,i_N,M_N\rangle \phantom{xxxx}
\cr
&\hphantom{xxxxxxxxxxx} \times\  \langle i_{\pi^{-1}(1)},M_{\pi^{-1}(1)}\,|\,
              i_{\sigma^{-1}(1)}',M_{\sigma^{-1}(1)}'\,\rangle\ \cdots\ 
\cr
&\hphantom{xxxxxxxxxxxxxxxxxx}
       \langle i_{\pi^{-1}(N)},M_{\pi^{-1}(N)}\,|\,
            i_{\sigma^{-1}(N)}',M_{\sigma^{-1}(N)}'\,\rangle 
\cr}\eq\sandwich$$
We have pulled out the conformal weights $h(i_k,M_k)$ of the states 
$|i_k,M_k\rangle$, and $c$ denotes the central charge of the component
theory. For the rhs to be non-vanishing, the representation labels and 
summation indices have to meet the following conditions, for 
all $k=1,\ldots,N\,$: 
\def\pms{\mathop{\hbox{${\scriptstyle \pi}\!\!$
\raise1.5pt\hbox{-${\scriptscriptstyle 1}$}${\scriptstyle \sigma}$}}}
$$\eqalign{
&i_k' = i_k\ ,\ \ i_k = i_{\pi(k)}\ ,\ \   i_k = i_{\sigma(k)}\ ,\ \   
i_k = i_{\pms(k)}\ ;
\cr
&M_k' = M_k\ ,\ M_k = M_{\pms(k)} \ .
\cr}\eq\sandwconds$$
The second line means that there are as many free summation indices 
in \sandwich\ as the permutation $\pi^{-1}\sigma$ has cycles -- 
namely $P^{{\ssstyle \pms}}$ 
in the above notations. For non-vanishing contributions to (10),
$M_k$ and $M_l$ are equal if $k$ and $l$ are in the same 
$(\pi^{-1}\sigma)$-cycle $C_\rho^{\pi^{-1}\sigma}$; 
as a consequence, the highest weight $h(i_k,M_k)$ appears 
with a factor $\Lambda_\rho\!\!\uupsi$. The $M_k$-summation then yields a 
$\cW$-character with argument $\tilde q{}^{\Lambda_\rho\!\!\upsi}$. 
After modular transformation to the open string channel, we end up with 
characters evaluated at fractional powers $q^{{1/\Lambda}}$.
\hbn
The restrictions on the representation labels $i_k$ are more severe 
since they contain the Ishibashi constraints from above. We can 
summarise them by inserting Kronecker symbols $\delta_{I,I'}$ and 
$\delta_I^{C^{\pi*\sigma}}$ which mean that the overlap vanishes 
unless $i_k = i_k'$ and  
$$
i_k = i_l \quad {\rm whenever}\ \ l = g(k) \ \ {\rm for\ some\ element}\ \ 
g \in \pi * \sigma :=
{\rm span}\{\pi,\sigma\} \subset S_N \ , 
$$ 
the subgroup of $S_N$ generated by $\pi$ and $\sigma$. 
One can show that $\delta_I^{C^{\pi*\sigma}}= \delta_I^{C^{\pi}} \! \cdot  
\delta_I^{C^{\sigma}}\;$ holds. (Below, we will sometimes refer to the 
orbits $C^{\pi*\sigma}$ of the subgroup $ \pi * \sigma$ as 
``cycles'', by slight abuse of terminology.)  With these abbreviations, 
and with the usual normalisation \q{\Ca} of Ishibashi states, we have 
$$\eqalign{
\hskip-5pt{}_\sigma\lla I'|\, \tqlc\,|I\rra_\pi &= 
\delta_{I,I'} \,\cdot\;\delta_I^{C^{\pi*\sigma}}\,\cdot\;
\chi_{{}_{i_{n_1}}} \bigl(\tilde q^{\Lambda_1}\bigr)\,\cdots 
\chi_{{}_{i_{n_P}}}\bigl(\tilde q^{\Lambda_{P}}\bigr)
\cr
&= \delta_{I,I'} \,\cdot\;\delta_I^{C^{\pi*\sigma}}\,\cdot\! 
  \sum_{j_1,\ldots,j_P}\ S_{i_{n_1} j_1} \cdots\, S_{i_{n_P} j_P}\ 
\chi_{{}_{j_1}} \bigl(q^{{1\over \Lambda_1}}\bigr)\,\cdots 
\chi_{{}_{j_P}}\bigl(q^{{1\over \Lambda_P}}\bigr)\ .    
\cr}
\eq\sandwresult$$
In a feeble attempt to avoid cluttering notations completely, we have 
omitted all superscripts ${}\upsi$ here, 
i.e.\ $n_\rho = n_\rho\!\!\uupsi$, $\Lambda_\rho = 
\Lambda_\rho\!\!\uupsi$, $P = P\uupsi$. Again, the $\chi_{{}_{j_\lambda}}$ 
denote characters of the original symmetry algebra $\cW$. Note that
the  cycles of $\pi^{-1}\sigma$ and $\sigma^{-1}\pi$ coincide up to 
internal reordering, therefore the result is symmetric in $\pi$ and 
$\sigma$. For $\pi \neq \sigma$, the characters in the last line 
of \sandwresult\ are familiar from the twisted sectors of cyclic 
orbifold theories. We will make further remarks on this at the 
end of the subsection. 
\def\hid{\raise5pt\hbox{${\scriptscriptstyle {\rm id}}$}}
\mn
Let us briefly look at two special cases before tackling Cardy's 
conditions in the general situation. For $\sigma = \pi$, the rhs of 
\sandwresult\ involves a product of $N= P\hid$ characters 
$\chi_{{}_{j_k}}(q)$ of $\cW$, and the cycle restrictions of $\pi*\pi$ 
are simply those already captured in $\delta^{C^\pi}_I$. Inserting the 
ansatz \ansatz\ for the full permutation boundary states, one finds 
(recall that $S = S^*$ with our assumptions) 
$$
Z_{\sfata_\pi\sfatb_\pi} (q) = \sum_{J=(j_1,\ldots,j_N)}\ 
 \Biggl\lb\; \prod_{\nu=1}^{P\upi}\  n_{\sfata_\pi\sfatb_\pi}^{(\nu)\;J}\; 
  \Biggr\rb\   \chi_{{}_{j_1}}(q)\cdots  \chi_{{}_{j_N}}(q)
\eq\genpfpipi$$\mkgho\genpfpipiho
with 
$$
n_{\sfata_\pi\sfatb_\pi}^{(\nu)\;J} = \sum_{i_{n_\nu}\in {\cal I}} \ 
  { S_{\alpha_\nu i_{n_\nu}} S_{\beta_\nu i_{n_\nu}} 
    \over  \bigl( S_{0\, i_{n_\nu}}\bigr)^{\Lambda_\nu}  }\; 
\prod_{k \in C^\pi_\nu}  S_{i_{n_\nu} j_k} \ \ .
$$
The numbers $n_{\sfata_\pi\sfatb_\pi}^{(\nu)\;J}$ can be calculated with 
the help of the quantum dimension property \qdimfus\ and of the Verlinde 
formula: 
$$\eqalignno{
n_{\sfata_\pi\sfatb_\pi}^{(\nu)\;J} &= \sum_{
               k_1,\ldots,k_{{}_{\Lambda_\nu-1}}}\! 
   N_{j_{n_\nu}\, j_{\pi(n_\nu)}}^{k_1} 
   N_{k_1\, j_{\pi^2(n_\nu)}}^{k_2} \!\cdots\;
   N_{k_{{}_{\Lambda_\nu-2}}\, j_{\pi^{\!\Lambda_{\!\nu}\!-\!1}(n_\nu)}}^{
                                         k_{{}_{\Lambda_\nu-1}}}\ 
   \sum_{i_{n_\nu}}\; { S_{\alpha_\nu\, i_{n_\nu}} S_{\beta_\nu\, i_{n_\nu}} 
    S_{k_{{}_{\Lambda_\nu-1}}\, i_{n_\nu}}  
   \over  S_{0\, i_{n_\nu}}  }
&\cr
&\qquad\ \ \ = \biggl(\  \prod_{k\in C^\pi_\nu} N_{j_k}\ \biggr)_{\alpha_\nu
                                            \beta_\nu}\ \ ,
&\mkg\npipi\cr}$$\mkgvo\npipivo
where we have used associativity of the fusion rules 
in the last step. 
All in all, we obtain the following partition function for two 
boundary states associated with the same permutation automorphism: 
$$
Z_{\sfata_\pi\sfatb_\pi} (q) = \sum_{j_1,\ldots,j_N}\ \; 
       \prod_{\nu=1}^{P\upi}\  
  \Biggl(\  \prod_{k\in C^\pi_\nu} N_{j_k}\ \Biggr)_{\!\!\alpha_\nu
                                            \beta_\nu}  
\;\chi_{{}_{j_1}}\!(q)\cdots  \chi_{{}_{j_N}}\!(q)\ \ .
\eq\pipf$$
This is a sum of $\cW^N$-characters with coefficients given by the 
$\alpha_\nu$-$\beta_\nu$-entries of ``cycle-wise'' products of fusion 
matrices $(N_i)_{jk} = N_{ij}^k$ from the component theory. Note 
that our permutation boundary states are ``orthonormal'' in the 
sense that the vacuum occurs (once) in the overlap \pipf\ 
if and only if$\fata_\pi = \fatb_\pi$. 
\sn
The case $\sigma = {\rm id} \neq \pi$ involves different products of 
$\cW$-characters, but is still easy to handle because 
$\delta_I^{C\,\upsi} = \delta_I^{C^{\pi*\sigma}} =
\delta_I^{C^{\pi}}$ holds. The boundary states $|\!|\fatb\rra_{\rm
id}$ are tensor products of Cardy boundary states for the
$\cW$-theory, but the $\delta_{I,I'}$-projection in the Ishibashi 
overlap \sandwresult\ means that only those $|I'\rra_{\rm id}$ 
contribute to the partition function that obey $i'_k = i'_{\pi(k)}$. 
Using this, eq.\ \sandwresult\ and our ansatz \ansatz, we get 
$$
Z_{\sfata_\pi\sfatb_{\rm id}} (q) = \sum_{j_1,\ldots,j_P} \ 
   \Biggl\lb\; \prod_{\nu=1}^P 
          n_{\sfata_\pi\sfatb_{\rm id}}^{(\nu)\;j_\nu}\;\Biggr\rb\; \ 
 \chi_{{}_{j_1}}\bigl(q^{{1\over \Lambda_1}}\bigr)\cdots  
 \chi_{{}_{j_P}}\bigl(q^{{1\over \Lambda_P}}\bigr)\ \,,
$$
with $P= P\uupi\;$, $\Lambda_\nu = \Lambda_\nu\!\!\!\!\uupi\;\;\,$, and 
with similar multiplicities as before:
$$ 
n_{\sfata_\pi\sfatb_{\rm id}}^{(\nu)\;j_\nu} = \sum_{i_{n_\nu}}\; 
  { S_{\alpha_\nu i_{n_\nu}} S_{i_{n_\nu} j_\nu} 
   \over   \bigl( S_{0\, i_{n_\nu}}\bigr)^{\Lambda_\nu}  }\; 
  \prod_{k \in C^\pi_\nu}  S_{\beta_k i_{n_\nu}} \ \ .
$$
As for the power in the denominator, the $\nu^{th}$ $\pi$-cycle
contributes $\Lambda_\nu/2$ and each of the $\Lambda_\nu$ id-''cycles'' 
contributes $1/2$. The $S$-matrix relations used above now yield the 
expression
$$
Z_{\sfata_\pi\sfatb_{\rm id}} (q) = \sum_{j_1,\ldots,j_P} \ 
    \Biggl\lb\  \prod_{\nu=1}^P\ 
         \biggl(\; \prod_{k \in C^\pi_\nu} N_{\beta_k} 
                        \biggr)_{\!\alpha_\nu j_\nu\,}\; \Biggr\rb \ 
\chi_{{}_{j_1}}\bigl(q^{{1\over \Lambda_1}}\bigr)\cdots  
\chi_{{}_{j_P}}\bigl(q^{{1\over \Lambda_P}}\bigr)\ \ .
\eq\idpipf$$
As a side-remark, we observe that the ``vacuum Cardy state'' 
$|\!|\hbox{\bf 0}\rra_{\rm id}$ of the tensor product theory 
provides a projection on single $\cW^{P\upi}$-characters in the 
sense that 
$$
Z_{\sfata_\pi \hbox{\smmfat 0}_{\rm id}} (q) =
\chi_{{}_{\alpha_1}}(q^{{1\over \Lambda_1}})\cdots \chi_{{}_{\alpha_{\!P\upi}}}
       (q^{{1\over \Lambda_{\!P\upi}}})\ \ .
\eq\idproj$$ 
Therefore, any boundary state $|\!| A\rra_\pi$ 
for the gluing conditions \permglu\ that is compatible with 
$|\!|\hbox{\bf 0}\rra_{\rm id}$ lies in the lattice cone spanned 
by the $|\!|\fata\rra_\pi$ . 
\sn
We have shown that $Z_{\sfata_\pi\sfatb_\sigma} (q)$ meets 
Cardy's conditions for $\sigma= \pi$ and for $\sigma = {\rm id}$. 
In the general case, the partition function is of the form 
$$
Z_{\sfata_\pi\sfatb_{\sigma}} (q) = \sum_{J=(j_1,\ldots,j_P)} \ 
\Biggl\lb\; \prod_{\lambda=1}^{P^*}\ 
   n^{(\lambda)\;J}_{\sfata_\pi\sfatb_{\sigma}}\; \Biggr\rb\ 
\chi_{{}_{j_1}}\bigl(q^{{1\over \Lambda_1}}\bigr)\cdots  
\chi_{{}_{j_P}}\bigl(q^{{1\over \Lambda_{P}}}\bigr)\ \ 
\eq\genpf$$
where now $P$ and the $\Lambda_\rho$ 
refer to number and lengths of the cycles of $\pi^{-1}\sigma$, 
while $P^*$ denotes the number of ``cycles'' or $C_\nu^{\pi*\sigma}$ 
of $\pi *\sigma$: The Kronecker symbols in \sandwresult\ imply that 
precisely $P^*$ of the $i_{n_\nu}$ are independent, while the 
constraints on the summation indices $M_k$ in \sandwconds\ leave us with
a product of $P$ characters. 
\hbn
The slightly lengthy proof for the fact that the coefficients 
$n^{(\lambda)\;J}_{\sfata_\pi\sfatb_\sigma}$ are indeed positive 
integers is given in Appendix A. 
All in all, we may conclude that the partition functions 
$Z_{\sfata_\pi\sfatb_{\sigma}} (q)$ satisfy Cardy's conditions for 
all permutations $\pi,\sigma \in S_N$ as long as the boundary state 
coefficients are given by formula \ansatz. Furthermore, any other 
(compatible) permutation boundary state lies in the integer lattice 
over these states. 
\mn
Some comments on the form \genpf\ of $Z_{\sfata_\pi\sfatb_\sigma}(q)$
are in order. These partition functions describe spectra of 
boundary fields, more specifically, if $\fata_\pi \neq 
\fatb_\sigma$, spectra of boundary condition changing operators 
(BCCOs). Already for the case $\pi=\sigma \neq {\rm id}$, the 
spectra in \genpfpipiho,\npipivo\ 
are in general different from those obtained with tensor products of 
component Cardy states, because of the cycle-wise products of fusion 
matrices in \npipi. 
\hbn
That the partition functions for $\pi \neq \sigma$ are built 
from characters $\chi_j(q^{1/\Lambda_\nu})$ can be understood 
as follows: While both gluing automorphisms $\Omega_\pi$ and 
$\Omega_\sigma$ preserve the full symmetry algebra $\cW^N$, it is 
a priori only the subalgebra $\cW^\circ$ with $\Omega_\pi({A}) = 
\Omega_\sigma({A})$ for all $A\in \cW^\circ$, i.e.\ the fixed-point 
algebra under $\Omega_{\pi^{-1}\sigma}$, that is represented on 
BCCOs which mediate
between the two gluing conditions. The form of this subalgebra 
is determined (up to isomorphism) by the cycle lengths $\Lambda_\nu$ 
of $\pi^{-1}\sigma$, i.e.\ by the conjugacy class of this 
permutation. We have $cW^\circ \simeq C_{\Lambda_1}\cW \times \cdots 
\times C_{\Lambda_P}\cW$ where $C_M\cW$ consists of all elements of 
$\cW^M$ that are invariant under the cyclic permutation of order $M$, 
i.e.\  $C_M\cW$ is the observable algebra of a cyclic 
$\Z_M$-orbifold, see \q{\KS}. This reference also shows how the characters 
$\chi_j\bigl(q^{{1\over M}}\bigr)$ enter the partition function of 
$\Z_M$ cyclic orbifolds: To construct such a partition function, one 
first projects the $M$-fold tensor product space onto invariant states; 
the states left fixed by $\Z_M$ contribute $Z(q^N,\,\bar q{}^N)$, where 
as before $Z$ belongs to the given component theory. To ensure modular 
invariance, one has to add twisted sectors, arising from these 
``fixed points'' after action with modular group generators \q{\KS}; 
this produces characters with fractional powers of $q$. One can express 
$$
\chi_j\bigl(q^{{1\over M}}\bigr) 
= \sum_{s=0}^{M-1} \ \chi_{(\widehat{j,s})}(q)
\eq\fracchartwistedchar$$ 
as a sum of cyclic orbifold characters $\chi_{(\widehat{j,s})}(q)$ 
corresponding to twisted sectors labelled by $s$; see \q{\BHS} for 
more details.
The highest weights on the rhs of \fracchartwistedchar\ are computed 
with $L_0$ from the cyclic orbifold model and read \q{\KS,\BHS} 
$$
h_{(j,s)} = { h_j+s \over M} + {M^2-1\over M}\,{c\over24}
\eq\corbhw$$ 
where $h_j$ and $c$ are conformal dimensions resp.\ central 
charge of the given component theory. 
\hbn
The decomposition \fracchartwistedchar\ now fits with general expectations: 
Computation of  open string partition functions 
$Z_{\sfata_\pi\sfatb_\sigma}(q)$ involves performing a modular transformation 
of traces of\  $\;\tilde q^{H^c} \Omega\dowpsi\,$, where $H^c$ is the 
closed string Hamiltonian. Indeed, this is one way to compute the partition 
function in the $(\pi^{-1}\sigma)$-twisted sector of cyclic orbifolds; see 
also \q{\MattTerr} for more details.
\sn
The symmetry algebra of a whole system $\fata^{i}_{\pi_i}$, $i=1,2,\ldots$, 
of permutation branes is the fixed point algebra 
$\bigl(\cW^N\bigr){}^{\cal F}$ with the subgroup 
${\cal F} = \langle\,\pi_i^{-1}\pi_j\,|\,i,j=1,\ldots\,\rangle \subset S_N$. 
Characters and fusion rules of such general permutation orbifolds have been 
studied in detail in Bantay's works \q{\Ban}. If all possible permutation
branes are included into the 
system, then ${\cal F} = S_N$ and we arrive at the symmetry algebra of 
a symmetric product. Thus, permutation branes have ``long'' open strings 
ending on them. In contrast to the setup in \q{\FM}, where boundary 
states for closed strings in a symmetric product background were 
constructed, here it is permutation gluing automorphisms that 
introduce long open strings into a theory of ordinary (``short'') 
closed strings. 
\bn
{\bf 2.4 Boundary OPE in a special case}
\mn
Beyond the cluster condition and Cardy's conditions, there are of course 
further sewing relations which have to be satisfied in a consistent 
boundary CFT \q{\Lew}. Most notably, the structure constants in the 
OPE of boundary fields are constrained by associativity. Solutions 
for the structure constants in terms of the fusing matrix have been 
worked out by Runkel \q{\Run} and later by other authors in \q{\BPPZ,\FFFS}. 
The formulae there apply to any unitary rational CFT (with diagonal 
bulk partition function) as long as Cardy-type boundary states for 
standard gluing conditions ($\Omega = {\rm id}$) are used. Thus,  
we can unfortunately not carry them over to our more complicated 
situation of permutation branes. We expect that a complete solution 
for the boundary OPE can be given once fusing matrices for permutation 
orbifolds are known. 
\hbn
For the time being, let us merely have a brief look at a simple special 
case, which already displays some interesting features. We restrict to 
a twofold tensor product with $\pi = (1\,2)$ and focus at the brane 
$|\!|\fata\rra_\pi = |\!|\hbox{\bf 0}\rra_\pi$. According to 
\genpfpipiho,\npipivo, the boundary spectrum for this special boundary 
condition is given by 
$$
Z_{\hbox{\smmfat 0}_\pi\, \hbox{\smmfat 0}_\pi}(q) = 
\sum_{j\in {\cal I}}\; \chi_{{}_j}(q)\,\chi_{{}_j}(q)\ \; .
\eq\zerozeropf$$
It ``coincides'' with that of the original bulk theory if we
``identify'' the right-moving charges of bulk fields with the 
second tensor factors of (chiral) boundary fields. (If we drop 
our assumption of self-conjugate sectors, \zerozeropf\ will be
replaced by the charge-conjugate bilinear and some of the following 
arguments have to be adapted accordingly.)
\hbn
This observation allows to solve the associativity condition for 
the OPE of boundary fields supported by $|\!|\hbox{\bf 0}\rra_\pi$:
That condition has the schematic form (for a self-conjugate theory)
$$
{\textstyle \sum_P}\; C_{IJP}\, C_{KLP}\; \hbox{{\sf F}}_{PQ}
    = C_{JKQ}\,C_{IQL}\ ;
$$
the $C$ are the OPE coefficients in question, indexed by $I=(i,i)$, 
$J=(j,j)$ etc.\ as counted in \zerozeropf, and {\sf F} is the fusing matrix
for the conformal blocks of the boundary CFT. In the special case at 
hand, the tensor product structure of the boundary fields implies that 
this fusing matrix is just the square of the fusing matrix of the 
component theory. But this means that the sewing relation for the 
boundary fields in \zerozeropf\ has the same form as the one for 
the bulk fields in the component theory. Therefore, the coefficients 
in the {\sl bulk} OPE of the original component theory provide a 
solution to the sewing relations of the {\sl boundary} OPE for the 
boundary condition $|\!|\hbox{\bf 0}\rra_\pi$ in the two-fold 
tensor product theory!
\hbn
We have not analysed sewing relations involving BCCOs yet, but 
assuming that the above OPE coefficients survive all further tests, 
one can in particular conclude that the boundary fields supported 
by $|\!|\hbox{\bf 0}\rra_\pi$ are mutually local. Namely, the bulk OPE 
coefficients ensure analyticity of the correlators. In view of the 
role of boundary OPEs for the analysis of non-commutative behaviour 
of branes \q{\ncg}, this suggests that the (low-energy) world-volumes 
of the branes $|\!|\hbox{\bf 0}\rra_\pi$ are commutative spaces, 
whatever the underlying component theory is. If a sigma-model 
interpretation of the original theory is available (with target 
$M$ say), then $|\!|\hbox{\bf 0}\rra_\pi$ should describe a 
brane whose world-volume is the diagonal in $M \times M$. This 
is confirmed by the results obtained in \q{\JoseSonia,\Antonetal} 
for WZW models with group target $G \times G$, using the classical 
interpretation of WZW gluing conditions resp.\ analysing the module 
structure of the algebra of boundary fields in the infinite level 
limit. 

\vfil\eject\noindent
{\bf 3. Permutation branes for the quintic}
\mn
Gepner models provide an important class of string backgrounds 
involving tensor products of rational CFTs \q{\Gep}. They were 
probably the first theories where cyclic permutation orbifolds 
of bulk CFTs were studied in the literature \q{\KS,\FKS}. In 
this context, the orbifold construction yields new closed string 
backgrounds, which correspond to new Calabi-Yau manifolds 
and to new spectra of massless closed (or heterotic) string states. 
Here, we want to use permutation gluing to obtain new (rational) 
branes for a given closed string  background, which remains unaltered. 
\mn
The partition functions of Gepner models are built from tensor product 
characters \q{\Gep}
$$
\chi^{\sfatl}_{\sfatm}(q) :=
  \chi_{s_0}(q)\; \chi^{l_1}_{m_1,s_1}(q)\, \cdots 
       \, \chi^{l_r}_{n_r,s_r}(q)
$$
where each $\chi^{l_j}_{m_j,s_j}(q)$ is a character of the 
$N=2$ super Virasoro algebra with level $k_j$ -- more precisely 
of the bosonic subalgebra (hence the additional label $s_j$). 
The $k_j$, $j=1,\ldots,r$,  are chosen in such a way that the 
central charges $c_j = 3k_j/(k_j+2)$ add up to 3, 6 or 9, 
corresponding to string compactifications down to $D=8$, 6 
resp.\ 4 external dimensions. $\chi_{s_0}(q)$ is a character 
of the $d = D-2$ free fermions associated with the transverse 
external directions. Our notations are as in \q{\Gep}, in 
particular $\fatm = (s_0; m_1,\ldots, m_r; s_1\ldots, s_r)$
with $s_0,s_j = 0,\ldots,3$, with $m_j= 0,\ldots, 2k_j+3$ and  
with  $l_j = 0, \ldots, k_j$. The combinations $l_j+m_j + s_j$ 
must be even. 
\hbn
Full Gepner model partition functions (those associated with 
SU(2) modular invariants of type $A$) have the form 
\def\sumlamube{\sum_{\hbox{{\sfatl},{\sfatm}}}\!{}^{{}^{\beta}} }
\def\sumlamubeR{\sum_{\hbox{{\sfatl},{\sfatm}}\in {\rm RR}}\!\!\!{}^{{}^{\beta}}\,}
\def\sumlamuevpr{\sum_{{\sfatl'},\,{\sfatm'}}\!{}^{{}^{\rm ev}} }

$$
Z_{\rm Gep}(q,\bar q) \sim 
\sum_{b_0,b_j} \sumlamube (-1)^{s_0} \;\chi^{{\sfatl}}_{{\sfatm}} (q)
\,\chi^{{\sfatl}}_{{\sfatm}+b_0\sfatb_0+b_1\sfatb_1+\ldots +b_r\sfatb_r}
(\bar q)
\eq\gepnerpartfct
$$
where $\fatb_0$ is the $(2r+1)$-vector with all entries  equal to 1, 
while  $\fatb_j$ has zeroes everywhere except for the first and the 
$(r+j+1)^{\rm st}$ entry which are equal to 2. The superscript ${}^\beta$ 
abbreviates Gepner's ``$\beta$-constraints'' in the summation, which 
implement the GSO projection and  ensure that all left-moving states 
are taken only from the NS sectors of the minimal models or only from 
the R sectors, see \q{\Gep}. We have $b_j = 0,1$ and $b_0 = 0,\ldots,K-1$ 
with $K:= {\rm lcm}(4,2k_j+4)$. 
\mn
After some earlier general considerations in \q{\OOY}, explicit boundary 
states for Gepner models were constructed in \q{\RSg}. The work \q{\BDLR} 
then introduced methods to relate abstract CFT boundary conditions to 
supersymmetric cycles in the Calabi-Yau regime, based on a computation 
of the intersection form at the Gepner point. This has triggered many 
interesting developments, see e.g.\ \q{\CY} for a very incomplete list 
of references, and has led to proposals for a new picture of branes 
(and bundles) on Calabi-Yau spaces \q{\DouCat}. 
\sn
All these investigations are based on Gepner model boundary states where 
the full tensor product symmetry algebra is preserved by imposing, in 
each of the $r$ minimal models individually, A- or B-type gluing 
conditions, which corresponds to choosing the gluing automorphism 
$\Omega^{(j)} = \Omega_{\rm mirror}$ resp.\ 
$\Omega^{(j)} = {\rm id}\,$\ for all $j=1,\ldots,r$. 
\hbn
Whenever some of the  levels $k_j$ coincide, a permutation group acts 
on the Gepner model, so one can look for permutation branes  -- which 
will still be rational and will still fall into the A- and B-type 
classification of \q{\OOY}, as the diagonal $N=2$ algebra is invariant 
under $\Omega_\pi$. Unfortunately, the prescription from Section 2 
cannot be used to construct new boundary states for Gepner models:  
The bulk partition functions \gepnerpartfct\ are neither diagonal 
nor charge-conjugate, due to the $\beta$-shifts in the right-moving 
charges, and therefore already the restrictions on representation 
indices, which select the available $\pi$-Ishibashi states, 
are different from those stated in 
\Ishiexcond\ above. Thus we have to adapt our methods to the Gepner case. 
\hbn
It will turn out that the set of admissible Ishibashi states depends
strongly on the relative divisibility properties of minimal model  
levels and cycle lengths of the permutation $\pi$. It appears, 
therefore, that permutation branes for Gepner models can only be 
constructed case by case, and we will more or less from the start
focus on the quintic $(3)^5$, which is the model studied in greatest 
detail in the literature. It has the added bonus that we need not 
worry about fixed point resolutions as discussed in \q{\BS,\FSW}. 
\hbn
In the next subsection, we will write down permutation boundary states 
for the quintic that satisfy A-type gluing conditions on the diagonal 
$N=2$ algebra, as well as compute the associated partition functions 
and intersection forms; then we will turn towards B-type permutation 
branes in Subsection 3.2. Formulae expressing some of the 
intersection forms in terms of charge symmetry generators are 
collected in Appendix B. 
\bn
{\bf 3.1 A-type boundary states}
\mn
Let us first determine which (A-type) Ishibashi states exist for a 
given permutation $\pi \in S_5$. 
An A-type  $\pi$-Ishibashi state for the quintic can be formed iff 
the left- and right-moving representation labels $(l_j,m_j,s_j)$ resp.\ 
$(l_j,m_j+b_0,s_j+b_0+2b_j)$ satisfy 
$$
 l_j = l_{\pi(j)}\ \ ,\quad
 m_j \equiv m_{\pi(j)} + b_0\ ({\rm mod}\;2k+4)\ \ ,\quad
 s_j \equiv s_{\pi(j)} + b_0 + 2b_{\pi(j)} \ ({\rm mod}\;4)
\eq\bbcyccond$$ 
for all $j=1,\ldots,5$ and for some choice of $b_0$ and $b_j$. 
In addition, since the external label $s_0$ is not affected by 
the permutation, A-type gluing requires that 
$$
  s_0 \equiv s_0 + b_0 + 2\sum_j b_j \ ({\rm mod}\;4)
\eq\socyccond$$
so that in particular $b_0$ must be even. The relations on the SU(2) 
labels $l_j$ in \bbcyccond\ are precisely as in the general RCFT 
setting discussed before -- and we anticipate that, in the SU(2) part, 
the same products of fusion matrices as in \pipf\ will show up in 
partition functions and intersection forms. All further complications 
arise from the additional summation indices $b_0,b_j$, which are 
constrained by the above equations, and by 
$$\eqalign{
&\Lambda_\nu b_0 \equiv 0\ \;({\rm mod}\;2k +4\,)\ ,
\cr
&\Lambda_\nu b_0 + 2\sum_{j\in C^\pi_\nu} b_j \equiv 0\ \;({\rm mod}\;4)
\cr}\eq\totcyccond$$
for all $\nu=1,\ldots,P^\pi$ simultaneously; the latter relations are 
obtained by applying \bbcyccond\ repeatedly until the $\pi$-cycle 
$C_\nu^\pi$ of length $\Lambda_\nu$ is closed. 
\hbn 
As soon as $\pi$ has a cycle of length one (i.e.\ a fixed point), the 
first condition in \totcyccond\ only leaves the possibilities $b_0=0$ 
or $b_0=2k+4$ so that, because of the label periodicities, there is just 
one independent $m$-value per $\pi$-cycle. 
The same is true for certain other permutations $\pi$ and certain 
other levels, e.g., if we specialise to $k=3$, for $\pi$ with 
$P^\pi=2$ and $\Lambda_1 =2$, $\Lambda_2=3$. For permutations in the 
conjugacy class of $\pi = (1\,2\,3\,4\,5)$, on the other hand, the 
first condition in \totcyccond\ admits all even $b_0 =0,2,\ldots,18$.  
\hbn 
As for the constraints on the summation variables $b_j$, it is not 
hard to see, by a case-by-case analysis for the quintic, that 
there is just enough freedom left to render all five $s_j$-labels 
independent (apart from the $\beta$-constraints). 
\sn
We summarise the free labels admitted by the Ishibashi constraints 
for permutation A-type gluing conditions in the case of the quintic. 
Although only the conjugacy class of $\pi$ will 
enter partition functions and intersection forms for two boundary 
states associated with the same gluing automorphism, it is 
more convenient to give a list for specific representatives:
\vskip.4cm
\setbox83=\vbox{$$xxxx \eq\AIshilist$$}
\vbox{
\halign{\ $\, # $\ \ \hfil & \rm # \hfill &#  \cr
\pi= {\rm id}\;{\rm :}
   &$I= (s_0;\; l_1,l_2,l_3,l_4,l_5;\; m_1,m_2,m_3,m_4,m_5;\;s_j)$ 
\cr
\pi= (1)(2)(3)(4\,5)\;{\rm :}
  &$I= (s_0;\;l_1,l_2,l_3,l_4,l_4;\; m_1,m_2,m_3,m_4,m_4;\;s_j)$
\cr
\pi = (1)(2\,3)(4\,5)\;{\rm :}
  &$I= (s_0;\; l_1,l_2,l_2,l_3,l_3;\; m_1,m_2,m_2,m_3,m_3;\;s_j)$ &
\cr
\pi = (1)(2)(3\,4\,5)\;{\rm :}
  &$I= (s_0;\; l_1,l_2,l_3,l_3,l_3;\; m_1,m_2,m_3,m_3,m_3;\;s_j)$ &
\cr 
\pi = (1)(2\,3\,4\,5)\;{\rm :}
  &$I= (s_0;\; l_1,l_2,l_2,l_2,l_2;\; m_1,m_2,m_2,m_2,m_2;\;s_j)$ &
\cr
\pi = (1\,2)(3\,4\,5)\;{\rm :}
  &$I= (s_0;\; l_1,l_1,l_2,l_2,l_2;\; m_1,m_1,m_2,m_2,m_2;\;s_j)$ &
\cr
\pi = (1\,2\,3\,4\,5)\;{\rm :}
  &$I= (s_0;\; l_1,l_1,l_1,l_1,l_1;\; m_1,m_1+2n,m_1+4n,m_1+6n,m_1+8n;
  \;s_j)$ &
\cr
&with $n=0,1,2,3,4$ &\hskip-25pt\AIshilist
\cr}}
\mn
To obtain an ansatz for the full boundary states, we combine the
expressions for Gepner branes from \q{\RSg} with formula \ansatz\ 
for permutation branes, and write
$$
|\!|\,\fata\,\rra_{A,\pi} \equiv 
   |\!|\,S_0; L_\nu, M_\nu, S_j \, \rra_{A,\pi} 
= {1\over\kappa^{A,\pi}_{\alpha}} \sumlamube
\;B^{{\sfatl} ,{\sfatm}}_{\sfata_{A,\pi}} \,
|{\fatl} ,{\fatm}\rangle\!\rangle_{A;\pi}\ \ .
\eq\Abdst$$
We have introduced one boundary state label per independent Ishibashi 
degree of freedom here: For $\pi$ with two or more cycles, we use one 
$L_\nu$-label and one $M_\nu$-label per $\pi$-cycle, along with labels 
$S_0$ and $S_j$ for $j=1,\ldots,5$. For the case $\pi = (1\,2\,3\,4\,5)$, 
one could start with five labels $M_j$, but it turns out that these boundary 
states depend only on the two quantities $M:=\sum_j M_j\;({\rm mod}\;10)$ and 
$M':=M_2+2M_3+3M_4+4M_5 \;({\rm mod}\;5)$. Note that the same $(M,M')$-labelling
also occurred in \q{\BDi} in connection with a $\Z_5$-orbifold of the 
quintic.  
\sn
The coefficients $B$ in \Abdst\ are given by 
$$
B^{{\sfatl} ,{\sfatm}}_{\sfata_{A,\pi}} =
(-1)^{{s_0^2\over2}} e^{-i\pi {s_0S_0\over2} }
\;e^{-{i\pi\over 2} \sum_{j=1}^5 s_jS_j }\ 
\Biggl\lb\;\prod_{\nu=1}^{P^\pi}\; 
{\sin \pi {(l_\nu+1)(L_\nu+1)\over 5} \over
\bigl(\sin \pi {l_\nu+1\over 5}\bigr)^{\Lambda_\nu/2}} \ 
e^{i\pi {m_\nu M_\nu\over 5} } \;\Biggr\rb
\eq\Apibdstcoeff
$$
if $\pi\in S_5$ has two or more cycles; for $\pi=(1\,2\,3\,4\,5)$,
we use the formula 
$$
B^{{\sfatl} ,{\sfatm}}_{\sfata_{A,\pi}} =
(-1)^{{s_0^2\over2}} e^{-i\pi {s_0S_0\over2} }
\;e^{-{i\pi\over 2} \sum_{j=1}^5 s_jS_j }\ 
{\sin \pi {(l_1+1)(L_1+1)\over 5} \over
\bigl(\sin \pi {l_1+1\over 5}\bigr)^{5/2}} \ 
e^{{i\pi\over 5} (m_1 M + 2n M')}
$$
where $n$ is the additional label showing up in the Ishibashi 
states for $\pi=(1\,2\,3\,4\,5)$. 
\sn
Due to the constraints on the 
$(l_j,m_j,s_j)$ allowed in Gepner models, the boundary 
state labels must satisfy $L_\nu+M_\nu+S_j \equiv 0\; ({\rm mod}\;2)\,$ 
for all $j\in C_\nu^\pi$. Furthermore, labels 
$(L_\nu,M_\nu,S_j)$ are identified 
with $(k-L_\nu,M_\nu+k+2,S_j+2)$ for all $j\in C_\nu^\pi$; for 
$S_5$-permutations with a single cycle, we identify $(L,M,M',S_j)$ 
with $(k-L,M+5,M',S_j+2)$. We restrict ourselves to even $S_j$. 
\mn
Now let us compute the partition functions 
$Z^A_{\sfata_\pi\,\tilde{\sfata}_\pi}(q)$ for two branes belonging 
to the same permutation. This is straightforward using Lagrange 
multipliers $\rho_0,\,\rho_j$ to disentangle the summations as in 
\q{\RSg}. We first introduce the abbreviation 
$$
f_{s,\rho} := (-1)^{s_0'+S_0-\tilde S_0} \ \sum_{\rho_j=0,1}\ 
\delta^{(4)}_{s_0'+  S_0-\tilde S_0 + \rho_0+ 2\; {\test 
            \Sigma_{\!{\scriptscriptstyle j}}} \rho_j+2}
\ \prod_{j=1}^5 \delta^{(4)}_{s_j'+S_j- \tilde S_j+\rho_0+2\rho_j}
$$
and, for simplicity, neglect the overall normalisation in the 
following; it can be fixed from Cardy's constraints as in 
\q{\RSg,\BDLR}. Then the results for the partition functions are 
$$\eqalignno{
Z^A_{\sfata_\pi\,\tilde{\sfata}_\pi}(q) &\sim 
\sumlamuevpr\ \sum_{\rho_0=0}^{19}\ 
f_{s,\rho} \ 
&\cr
&\qquad\quad \times \Biggl\lb\;\prod_{\nu=1}^{P^\pi}\;
   \Bigl(\,\prod_{n \in C_\nu^\pi} N_{l_n'}\Bigr)_{L_\nu\tilde L_\nu} \;
\delta^{(10)}_{{\textstyle \Sigma}_{{}_{\!{\scriptscriptstyle n \in
C_\nu^\pi}}} m_n'\  + M_\nu-\tilde M_\nu+\Lambda_\nu \rho_0}\,
 \Biggr\rb \ \chi^{{\sfatl'}}_{{\sfatm'}} (q)\ 
&\mkg\ZpipiGepn
\cr}$$
for $P^\pi >1$. The superscript ${}^{\rm ev}$ indicates that  
$l_j' + m_j' + s_j'$ must be even. For the permutation 
$\pi = (1\,2\,3\,4\,5)$, the Kronecker symbol in 
\ZpipiGepn\ is to be replaced by 
$$
\delta^{(10)}_{m_1'+\ldots +m_5' + M-\tilde M + 5\rho_0}\ 
 \delta^{(5)}_{m_2'+2 m_3'+ 3m_4' + 4m_5' + M'-\tilde M'}\ \ . 
\eq\deltarepl$$
Comparing to the partition functions for ordinary ``un-permuted''  
A-type branes \q{\RSg}, we observe that now products of fusion 
matrices occur, as in the diagonal RCFT case discussed in Section 2. 
Moreover, the Kronecker restrictions on the labels $m_j'$ of the 
characters $\chi^{{\sfatl'}}_{{\sfatm'}} (q)$ are different from
those for $\pi = {\rm id}$. By spelling out the partition functions 
explicitly, one can check that the permutation branes 
in particular support new spectra of massless open string states. 
On the other hand, the expressions \ZpipiGepn\ enjoy stability 
and space-time supersymmetry just as those for $\pi= {\rm id}$. 
\mn
In the intersection form $I^A_{\sfata_\pi\,\tilde{\sfata}_\pi}$, 
similar building blocks as in \ZpipiGepn\ show up. We use the 
definition \q{\DFi,\BDLR}
$$
I^A_{\sfata_\pi\,\tilde{\sfata}_\pi}
  = {\rm tr}_{{\cal H}_{\rm R}} (-1)^F q^{L_0-{c\over24}}
= \sumlamubeR B^{{\sfatl} ,{\sfatm}}_{\sfata_\pi} 
              B^{{\sfatl} ,{-\sfatm}}_{\sfata_\pi}
 {}_{A;\pi}\lla \fatl,\fatm |\,(-1)^{F_L} \tqlc\,|\fatl,\fatm\rra_{A;\pi}
$$
with $(-1)^{F_L} := (-1)^{J_0^{\rm int}+ {d'\over2}}$, where 
$J_0^{\rm int}$ is the charge in the internal sector and $d' := 4
-{d\over2} = \oh\,d^{\rm int}$. Trace and summation run over 
Ramond sectors only. 
Proceeding along the lines of \q{\BDLR} -- exploiting  
$\beta$-constraints, field identifications and the fact that 
only R ground states (i.e.\ primaries with labels $(l_j',l_j'+1,1)$ 
or $(l_j',-l_j'-1,-1)$) contribute to this index --  
we arrive at 
$$
I^A_{\sfata_\pi\,\tilde{\sfata}_\pi} \sim 
\sum^9_{{m_1',\ldots,m_5'};\;\rho_0 =0}\  
\Biggl\lb\;\prod_{\nu=1}^{P^\pi}\;
  \Bigl(\,\prod_{n \in C_\nu^\pi} N_{m'_n-1}\Bigr)_{L_\nu\tilde L_\nu}\;
\delta^{(10)}_{ {\textstyle \Sigma}_{{}_{\!
      {\scriptscriptstyle n \in C_\nu^\pi}}} m_n'\ + 
          M_\nu-\tilde M_\nu+\Lambda_\nu (2\rho_0+1)}\; \Biggr\rb \  .
\phantom{x}\eq\intpipi
$$
The intersection form for the one-cycle case is obtained by 
the same replacement of the Kronecker symbol as in \ZpipiGepn\ 
and \deltarepl.
\hbn
In order to save signs in \intpipi, we have used the extension 
$N^{-l-2}_{L\tilde L} = -N^{l}_{L\tilde L}$ and $N^{-1}_{L\tilde L}
= N^{k+1}_{L\tilde L}=0$ of the SU(2) fusion rules to ``spins'' $l$  
beyond $0,\ldots,k$, see \q{\BDLR}; analogously if two or three 
spins are in the extended range. 
\mn
As we have seen already for diagonal RCFTs in Subsection 2.3,  partition 
functions $Z^A_{\sfata_\pi\,\tilde{\sfata}_\sigma}(q)$ for two different
permutations depend very much on the relative ``location'' of the 
cycles of $\pi$ and $\sigma$. We could compute such partition 
functions and intersection forms in a relatively straightforward 
fashion using the formulae from above, but we restrict ourselves to 
the special case $\sigma = {\rm id}\,$ here.
\hbn
When evaluating overlaps between $\pi$- and ${\rm id}$-Ishibashi 
states, one observes (similarly to the previous general considerations)
that further constraints on the labels arise. Obviously, the
(independent) $l_j$ and $m_j$ labels from the ${\rm id}$-states 
have to match their (constrained) partners from the $\pi$-states.
Moreover, the overlaps are non-zero only for 
specific choices of $s_j$ labels (this follows as in \sandwconds), 
although existence of Ishibashi states alone imposes no relations 
on these labels. All in all, one is 
left with precisely one independent (except for $\beta$-constraints) 
array $(l_\nu,m_\nu,s_\nu)$ per $\pi$-cycle to sum over. 
\hbn
Writing down partition functions and intersection forms is 
therefore rather easy. For the spectra of boundary condition 
changing operators between quintic A-type branes we obtain  
\vbox{
$$\eqalignno{
Z^A_{\sfata_\pi\,\tilde{\sfata}_{\rm id}}(q) &\sim 
\sumlamuevpr\ \sum_{\rho_0=0}^{19}\ \sum_{\rho_\nu=0,1}\ 
 (-1)^{s_0'+S_0-\tilde S_0} \
\delta^{(4)}_{s_0'+  S_0-\tilde S_0 + \rho_0+ 2\, {\test 
            \Sigma_{\!{\scriptscriptstyle \nu}}} \,\rho_\nu\ +2}
\phantom{xxxxx}&\cr
&\qquad\quad \times \Biggl\lb\ \prod_{\nu=1}^{P^\pi}\;
   \Bigl(\,\prod_{j \in C_\nu^\pi} N_{\tilde L_j}\Bigr)_{L_\nu l_\nu'}\;
\delta^{(10)}_{m_\nu' +  M_\nu - 
 {\textstyle \Sigma}_{\!{\scriptscriptstyle j \in C_\nu^\pi}}\tilde M_j \ 
  +\Lambda_\nu \rho_0}\;
&\cr
&\phantom{MMMMMMMM}\times\ \delta^{(4)}_{s_\nu' +  
 {\textstyle \Sigma}_{\!{\scriptscriptstyle j \in C_\nu^\pi}}
                                   (S_j -\tilde S_j)\ 
  +\Lambda_\nu \rho_0 + 2\rho_\nu}\;
\Biggr\rb \ {}^{[\pi]}\chi^{\sfatl'}_{{\sfatm'}} (q)
&\mkg\ZpiidGepn
\cr}$$
}
with the expected combinations of cyclic orbifold characters 
$$
{}^{[\pi]}\chi^{\sfatl'}_{{\sfatm'}} (q) = 
\chi_{{}_{s_0}}(q)\; \chi_{{}_{(l_1',m_1',s_1')}}
  \bigl(q^{1\over \Lambda_1}\bigr) \cdots  
    \chi_{{}_{(l_P',m_P',s_P')}}\bigl(q^{1\over \Lambda_P}\bigr)\ \ .
$$
Let us have a closer look at the $\delta^{(4)}$-constraint in the last line 
of \ZpiidGepn: It prevents characters with $s_\nu'$ odd from contributing 
to the partition function whenever $\pi$ has a cycle $C^\pi_\nu$ 
of even length $\Lambda_\nu$. This does not mean, however, that strings 
stretching between an ordinary and a $\pi$-brane for such a $\pi$ do 
not have a Ramond sector. One has to recall\footnote{${}^1$}{I 
am indebted to Matthias Gaberdiel for a crucial discussion of this point.} that 
the modes of $\Z_{\Lambda_\nu}$-twisted R-generators are shifted by 
integer multiples of $1/\Lambda_\nu$, so that for even $\Lambda_\nu$ 
the minimal model characters with even $s_\nu'$ may actually belong to 
twisted R-representations. 
\sn
The same effect has to be taken into account when computing the 
intersection form between a $\pi$-brane and an ordinary A-type brane. 
The massless states (in the space-time sense) that contribute to 
$I^A_{\sfata_\pi\,\tilde{\sfata}_{\rm id}}$ are tensor products 
of ordinary minimal model Ramond ground states for the cycles of 
odd length, and states with 
$$ 
{1\over \Lambda_\nu}\,h_{(l_\nu',m_\nu',s_\nu')} = {\Lambda_\nu\, c\over 24}
$$
for cycles $C^\pi_\nu$ with $\Lambda_\nu$ even, where $c={3k\over k+2}$. For the 
quintic, these states are labelled by 
$$\eqalign{
&\Lambda_\nu = 2\;:\ \ (l_\nu',m_\nu',s_\nu') = (3,\pm 3,0) 
\cr
&\Lambda_\nu = 4\;:\ \ (l_\nu',m_\nu',s_\nu') = (3,\pm 1,0) 
\cr}\eq\twistedmassless$$
up to field identification. To verify this, one has to go beyond the usual 
$h\,$(mod 1) expressions given in the literature and work with the true conformal 
dimensions of $N=2$ minimal model primaries, which can be obtained from the coset 
construction.\footnote{$\!{}^2$}{I thank Stefan Fredenhagen for making his 
private notes available to me.} The intersection form 
$I^A_{\sfata_\pi\,\tilde{\sfata}_{\rm id}}$ is a product of one term 
$$
 \Bigl(\,\prod_{j \in C_\nu^\pi} N_{\tilde L_j}\Bigr)_{L_\nu,m_\nu'-1}\ 
\delta^{(10)}_{m_\nu' +  M_\nu - 
 {\textstyle \Sigma}_{\!{\scriptscriptstyle j \in C_\nu^\pi}}\tilde M_j
  \ +\Lambda_\nu (2\rho_0+1)}
$$
per odd length cycle (with $\rho_0=0,\ldots,9$) and, for each even length 
cycle, a term where $m_\nu'-1$ resp.\ $m_\nu'$ are replaced by the $l_\nu'$ 
resp.\ $m_\nu'$ values from \twistedmassless.

\bn
{\bf 3.2 B-type boundary states}
\mn
Along the same lines as above, one can determine $\pi$-permuted B-type 
boundary states for Gepner models. The A-type Ishibashi conditions 
\bbcyccond\ and \socyccond\ are replaced by 
$$ 
l_j = l_{\pi(j)}\ \ ,\quad
 - m_j \equiv m_{\pi(j)} + b_0\ ({\rm mod}\;2k+4)\ \ ,\quad
 - s_j \equiv s_{\pi(j)} + b_0 + 2b_{\pi(j)} \ ({\rm mod}\;4)
\eq\bbcyccondB$$ 
and 
$$
  - s_0 \equiv s_0 + b_0 + 2\sum_j b_j \ ({\rm mod}\;4)\ \ .
\eq\socyccondB$$
The condition on the $m_j$ implies $m_{\pi^l(j)} = m_{\pi^{l+2}(j)}$
for all $l$, thus there are at most two independent $m_j$-values 
per $\pi$-cycle; more precisely 
$$\eqalign{
2 m_j &\equiv -b_0 \ ({\rm mod}\;2k+4)\hskip45pt {\rm for\ all}\ 
   j \in C_\nu^\pi\ \ {\rm if}\ \,\Lambda_\nu\ {\rm is\ odd}\  ,
\cr
m_{\pi(j)} &\equiv - m_{j} - b_0\ ({\rm mod}\;2k+4)\quad \quad 
     {\rm for\ all}\ 
   j \in C_\nu^\pi\ \ {\rm if}\ \,\Lambda_\nu\ {\rm is\ even} \ .
\cr}$$
Similarly, the constraints on the $s$-labels require that $b_0$ is even 
and that 
$$\eqalign{
2\sum_{n\in  C_\nu\upi} b_n &\equiv 0 \ ({\rm mod}\;4)
\hskip31pt  {\rm if}\ \,\Lambda_\nu\ {\rm even}\ \; ,
\cr
2s_j + 2\sum_{n\in  C_\nu\upi} b_n &\equiv -b_0 \ ({\rm mod}\;4)\quad 
\quad{\rm for\ all}\ j \in C_\nu^\pi\quad 
     {\rm if}\ \,\Lambda_\nu\ {\rm odd} \ \; .
\cr
}$$
\sn
It is straightforward to work out a list of admissible permuted 
B-type Ishibashi states for the $(3)^5$ model. As before, $s_0$ 
and the five $s_j$-values are only restricted by the 
$\beta$-constraints, and we have the following label structure:
\vskip.3cm
\setbox84=\vbox{$$xxxx \eq\BIshilist$$}\def\om{\overline{m}}
\vbox{
\halign{\ $\, # $\ \ \hfil & \rm # \hfill &#  \cr
\pi= {\rm id}\;{\rm :}
   &$I= {\scst (s_0;\; l_1,l_2,l_3,l_4,l_5;\; -b_0'+ 5 \alpha_1,\,
                   -b_0'+ 5 \alpha_2,\,-b_0'+ 5 \alpha_3,\,
                   -b_0'+ 5 \alpha_4,\,-b_0'+ 5 \alpha_5;\;s_j)}$& 
\cr
\pi = (1)(2)(3)(4\,5)\;{\rm :}
  &$I= {\scst (s_0;\;l_1,l_2,l_3,l_4,l_4;\; -b_0'+ 5 \alpha_1,\,
                   -b_0'+ 5 \alpha_2,\,-b_0'+ 5 \alpha_3,\,
                   \overline{m}_4,\,-\om_4 - 2 b_0';\;s_j)}$&
\cr
\pi = (1)(2\,3)(4\,5)\;{\rm :}
  &$I= {\scst (s_0;\; l_1,l_2,l_2,l_3,l_3;\;  -b_0'+ 5 \alpha_1,\,
   \om_2,\,-\om_2-2b_0',\,\om_3,\,-\om_3-2b_0';\;s_j)}$&
\cr
\pi = (1)(2)(3\,4\,5)\;{\rm :}
  &$I= {\scst (s_0;\; l_1,l_2,l_3,l_3,l_3;\; -b_0'+ 5 \alpha_,1,\,
                   -b_0'+ 5 \alpha_2,\,-b_0'+ 5 \alpha_3,\,
                   -b_0'+ 5 \alpha_3,\,-b_0'+ 5 \alpha_3;\;s_j)}$
&
\cr 
\pi = (1)(2\,3\,4\,5)\;{\rm :}
  &$I= {\scst (s_0;\; l_1,l_2,l_2,l_2,l_2;\; -b_0'+ 5 \alpha_1,\,
     \om_2,\,-\om_2-2b_0',\,\om_2,\,-\om_2-2b_0';\;s_j)}$&
\cr
\pi = (1\,2)(3\,4\,5)\;{\rm :}
  &$I= {\scst (s_0;\; l_1,l_1,l_2,l_2,l_2;\; \;\om_1,\,-\om_1-2b_0',\,
                 -b_0'+ 5 \alpha_1,\,
                   -b_0'+ 5 \alpha_1,\,-b_0'+ 5 \alpha_1;\;s_j)}$ &
\cr
\pi = (1\,2\,3\,4\,5)\;{\rm :}
  &$I= {\scst (s_0;\; l_1,l_1,l_1,l_1,l_1;\; -b_0'+ 5 \alpha_1,\, 
      -b_0'+ 5 \alpha_1,\, -b_0'+ 5 \alpha_1,\, -b_0'+ 5 \alpha_1,\,
               -b_0'+ 5 \alpha_1;\;s_j)}$&\BIshilist 
\cr}}
\mn
Here, the $\om_i$ and $b_0'$ range over $0,1,\ldots,9$, while $\alpha_i = 0,1$. 
\mn
Permutation B-type boundary states can be constructed with coefficients 
very similar to those in \Apibdstcoeff: 
$$
B^{{\sfatl} ,{\sfatm}}_{\sfata_{A,\pi}} =
(-1)^{{s_0^2\over2}} e^{-i\pi {s_0S_0\over2} }
\;e^{-{i\pi\over 2} \sum_{j=1}^5 s_jS_j }\ 
\Biggl\lb\; \prod_{\nu=1}^{P^\pi}\; 
{\sin \pi {(l_\nu+1)(L_\nu+1)\over 5} \over
\bigl(\sin \pi {l_\nu+1\over 5}\bigr)^{\Lambda_\nu/2}} \,\Biggr\rb \ 
\prod_{j=1}^{5}\; e^{i\pi {m_j M_j\over 5} } 
\eq\Apibdstcoeff
$$
where the $m_j$ are to be expressed by $b_0$, $\alpha_i$ and $\om_i$ 
as in the list above. The labels $S_0,S_j$ and $L_\nu$, $\nu=1,
\dots,P^\pi$, are as for A-type gluing conditions. 
Closer inspection shows that B-type boundary states  
depend only on $L_\nu$ and $S_j$ and the following 
combinations of the five $M_j$:
$$\eqalign{
M &:= \sum_{j=1}^5  M_j\ \ ({\rm mod}\;10)\ \;,
\cr
M_{[\nu]} &:= M_{n\upi_\nu} -  M_{\pi(n\upi_\nu)} + - 
  \ldots - M_{\pi^{\Lambda_\nu -1}(n\upi_\nu)} \ \ ({\rm mod}\;10)
\quad\ \hbox{\rm if\ $\Lambda_\nu$\ even}\ \ ;
\cr}\eq\indepM$$
here, $n\upi_\nu$ denotes a chosen representative of the 
cycle $C_\nu^\pi$ as in \picycles. Thus, $\pi$-permuted 
B-type boundary states with $\pi$ from the conjugacy classes
in the first, fourth and last row of \BIshilist\ are 
distinguished by the single label $M$ (together with $L_\nu$ 
and $S_j$ of course), while for the other conjugacy classes one 
also has to specify the values of the alternating 
$M_j$-sums over even length cycles. We 
have the constraints $L_\nu+M_j+S_j =0\; ({\rm mod}\;2)$ for all 
$j\in C_\nu^\pi$, as well as the identification 
$(L_\nu,M,M_{[\nu]},S_j) \equiv (k-L_\nu,M+5,M_{[\nu]}+5,S_j+2)$. 
\sn
The label structure of permuted B-type branes, which differs from 
what occurred in the A-type cases, is also reflected in the formulae 
for B-type partition functions and intersection forms. One finds the 
following result: 
$$\eqalignno{
Z^B_{\sfata_\pi\,\tilde{\sfata}_\pi}(q) &\sim 
\sumlamuevpr\ \sum_{\rho_0=0}^{19}\ f_{s,\rho} \ 
         \delta^{(10)}_{5 \rho_0 + M - \tilde M + 
                {\textstyle \Sigma}_{{}_{\!{
                 \scriptscriptstyle j}}} m_j'}
\ \;\Biggl\lb\; \prod_{\nu=1}^{P^\pi}\;
   \Bigl(\,\prod_{n \in C_\nu^\pi} N_{l_n'}\Bigr)_{
                                L_\nu\tilde L_\nu}\, \Biggr\rb \ 
\phantom{xxxxxxxxxxxx}&\cr
&\phantom{xxx}\times\ \Biggl\lb\;
     \prod_{\nu=1 \atop \Lambda_\nu\ {\rm odd}}^{P^\pi}\;
  \delta^{(2)}_{\rho_0+ M_\nu- \tilde M_\nu+ {\textstyle \Sigma}_{{}_{\!{
     \scriptscriptstyle n \in C_\nu^\pi}}} m_n'} \,\Biggr\rb\ 
\Biggl\lb\;\prod_{\nu=1 \atop \Lambda_\nu\ {\rm even} }^{P^\pi}\;
\delta^{(10)}_{\, M_{[\nu]} - \tilde M_{[\nu]} + m_{[\nu]}'}\, \Biggr\rb \ 
  \chi^{{\sfatl'}}_{{\sfatm'}} (q)\ 
&\mkg\ZpipiGepnB
\cr}$$
where we have used the abbreviation $m_{[\nu]}'$ for the alternating 
sum of $m_n'$ over even cycles in analogy to $M_{[\nu]}$ in \indepM.
The summation in the first Kronecker symbol runs over all $j =1,\ldots,5$.
\sn
The intersection form for B-type boundary conditions $\fata_\pi$ and 
$\tilde{\hbox{$\fata$}}_\pi$ associated with the same permutation $\pi$ reads
$$I^B_{\sfata_\pi\,\tilde{\sfata}_\pi} 
\sim 
\sum_{m_1',\ldots,m_5'}\ 
         \delta^{(10)}_{5 + M - \tilde M + 
                {\textstyle \Sigma}_{{}_{\!{
                 \scriptscriptstyle j}}} m_j'}
\ \prod_{\nu=1}^{P^\pi}\;
   \Bigl(\,\prod_{n \in C_\nu^\pi} N_{m_n'-1}\Bigr)_{
                                L_\nu\tilde L_\nu}\, 
\prod_{\nu=1 \atop \Lambda_\nu\ {\rm even} }^{P^\pi}
\delta^{(10)}_{\, M_{[\nu]} - \tilde M_{[\nu]} + m_{[\nu]}'}\  .
\phantom{x}\eq\intpipiB$$
The summation index $\rho_0$ would appear only via the combination 
$5(2\rho_0+1)$ in the $\delta^{(10)}$, thus drops out. 
\hbn
As before, one can compute partition functions and intersection 
forms between un-permuted and $\pi$-permuted B-type branes, and the 
observations on even cycle length $\Lambda_\nu$ from Subsection 3.1 again apply. 

\bn\mn
{\bf 4. Open problems}
\mn
We have considered tensor products of rational CFTs and studied boundary 
conditions governed by gluing automorphisms from the permutation 
group. We have presented an ansatz for the associated permutation 
boundary states and checked that cluster and Cardy's conditions 
are satisfied. We were able to write down explicit expressions for  
the open string spectra $Z_{\sfata_\pi\sfatb_\sigma} (q)$ for 
the cases $\pi = \sigma$ and $\pi \neq \sigma = {\rm id}$. As 
a new feature compared to tensor products of ordinary Cardy 
branes, cycle-wise products of fusion matrices show up in 
these partition functions. By making use of decompositions of 
permutations more cleverly, it should be possible to go beyond 
the non-constructive integrality proof of Appendix A and find closed 
formulae for the multiplicities 
$n^{(\lambda)\;J}_{\sfata_\pi\sfatb_\sigma}$ in the partition functions 
for arbitrary $\pi$ and $\sigma$. 
\hbn
Whenever $\pi \neq\sigma$, the partition function involves characters 
of twisted representations. Such characters also play a major role 
in the recent work \q{\MattTerr}, where twisted boundary conditions 
for WZW models were studied; as in the present paper, the gluing 
conditions are not ``aligned'' with the automorphism that determines 
the bulk partition function. In \q{\MattTerr}, the multiplicities 
in the open string partition function were expressed in terms of 
S-matrix elements of ordinary and twined characters, and it would 
be interesting to compare these expressions to the formulae given 
here, in the cases covered by both points of view; the results of 
\q{\Ban} should prove useful in the process. (A special example 
occurs in \q{\MattSaku}, where branes in an asymmetric torus 
orbifold connected with the 5-fold tensor product of 
$\widehat{\rm su}(2)_1$ are studied. The authors could in 
particular show that the prescriptions given here and in 
\q{\MattTerr}, suitably adapted to the non-diagonal partition 
function of this model, indeed lead to the same boundary states.) 
\hbn
We have not been able to say too much about boundary OPEs and the 
associativity constraints they must satisfy. Investigating the 
relation of conformal blocks in cyclic orbifolds to those from the 
component theory should be relevant here. The one, very simple brane 
for which we could study the boundary OPE without high-powered 
techniques revealed an intriguing connection between (component 
theory) closed string interactions and interactions of (tensor 
product) brane excitations. It would be interesting to relate 
this to the findings of \q{\PZ}, where new connections between 
boundary conditions and structures in the bulk were uncovered. 
\sn 
In the second part of the paper, we have presented permutation 
branes for the quintic at the Gepner point. We have given 
formulae for open string spectra and intersection forms and 
prepared the ground for a geometric interpretation of the 
new branes by providing expressions of for some 
$I_{\sfata_\pi\,\tilde{\sfata}_\pi}$ in terms of ``quantum symmetry'' 
generators. Obviously, such expressions should be found for the 
missing cases (large cycle lengths and $\pi \neq \sigma$), and one 
should systematically compute topological invariants of the 
associated bundles (for B-type branes), following the lines 
of \q{\BDLR,\CY}. 
\hbn
But even without a detailed analysis, 
the intersection forms for B-type permutation branes listed in 
Appendix B suggest that among the boundary states for 
$\pi = (1)\,(2)\,(3)\,(4\,5)$ there is one with the charges 
of a configuration made up from D-branes only. (Note that 
this does not contradict the arguments given in \q{\Scheid}, 
which assume $\pi={\rm id}$.) The charges of some of the new 
branes will be sums of charges of the ``old'' boundary states 
from \q{\RSg}, suggesting that they can be seen as bound states. 
This might provide tests for some of the conjectures arising 
from the derived category picture of B-type Calabi-Yau branes 
developed by Douglas \q{\DouCat}.
\hbn
The A-type permutation branes, on the other hand, can perhaps 
be exploited for a construction of new special Lagrangian cycles 
for the quintic, using the linear sigma model  methods uncovered 
in \q{\GJS,\HIV} and developed further in \q{\Kenn}. 

\bn\sn
{\bf Acknowledgements}\quad For discussions, comments, encouragement, 
I am grateful to M.\ Douglas, M.\ Gaberdiel, C.\ R\"omelsberger, to 
C.\ Bachas, P.\ Bantay, I.\ Brunner, S.\ Fredenhagen, K.\ Graham, 
S.\ Gukov, K.\ Kennaway, G.\ Papadopoulos,  D.\ Roggenkamp, F.\ Roose, 
S.\ Sch\"afer-Nameki, E.\ Scheidegger, V.\ Schomerus, G.\ Watts, 
K.\ Wendland, E.\ Zaslow, J.-B.\ Zuber, and to Sonia Stanciu. 
\hbn
Hospitality by ENS, Paris, and ESI, Vienna, during various stages of 
this project is gratefully acknowledged. This work was supported in part 
by the Nuffield Foundation, Grant Number NUF-NAL/00421/G, by the PPARC 
grant PPA/G/S/1998/0061 and by the EU Superstrings Network HPRN-CT-2000-00122.

\vfil\eject\noindent
{\bf Appendix A}
\mn
Here we fill the gap left in Subsection 2.3 and prove that, in 
a tensor product of diagonal rational CFTs, the 
partition functions  $Z_{\sfata_\pi\sfatb_{\sigma}}(q)$ obey Cardy's 
conditions for arbitrary permutations $\pi,\sigma \in S_N$. This means 
we have to show that the coefficients 
$$
n^{(\lambda)\;J}_{\sfata_\pi\sfatb_\sigma} = 
\sum_{i_{n^*_\lambda} \in\,{\cal I}}\ \;
    \prod_{n\upi_\nu\in\, C^{\pi*\sigma}_\lambda} 
              B_{\alpha_\nu i_{n^*_\lambda}}\;
    \prod_{n\usi_\mu\in\, C^{\pi*\sigma}_\lambda} 
              B_{\beta_\mu i_{n^*_\lambda}}\;
    \prod_{n\upsi_\rho\in\, C^{\pi*\sigma}_\lambda} 
              S_{j_{n_\rho}i_{n^*_\lambda}}
\eq\genn$$
from eq.\ \genpf\ are positive integers. Here, 
$n^*_\lambda$ is a fixed representative of the $\lambda^{\rm th}$ orbit 
$C^{\pi*\sigma}_\lambda$ of $\pi*\sigma$, and for the 
representatives $n\upi_\nu$, $n\usi_\mu$ and $n_\rho \equiv 
n\upsi_\rho$ of $\pi$-, $\sigma$- resp.\ $(\pi^{-1}\sigma)$-cycles 
which intersect $C^{\pi*\sigma}_\lambda$, we have put $i_{n\upi_\nu} 
= i_{n^*_\lambda}$ etc., implementing the  Kronecker symbols in 
\sandwresult. (Note that, since we fixed the representatives 
$n^\pi_\nu$ etc.\ once and for all, the first product simply runs 
over all $\pi$-cycles that are contained in the $\lambda^{\rm th}$
orbit $C^{\pi*\sigma}_\lambda$ of the group $\pi * \sigma$.) 
\hbn
The coefficients $B$ in \genn\ contain half-integer powers of 
$S_{0\, i_{n^*_\lambda}}$ in the denominator. {}From the definition 
of $C^{\pi*\sigma}_\lambda$, it is however clear that 
$$
\sum_{\nu:\; n\upi_\nu\in\, C^{\pi*\sigma}_\lambda}\!\! \Lambda\uupi_\nu\ \ = 
\sum_{\mu:\; n\usi_\mu\in\, C^{\pi*\sigma}_\lambda}\!\! \Lambda\uusi_\mu\ \ = 
\  \Lambda_\lambda^*\ \ ,
$$
the cardinality of $C^{\pi*\sigma}_\lambda$. Thus, the denominator 
in \genn\ is simply $\bigl(S_{0\, i_{n^*_\lambda}}\bigr)^{
\Lambda^*_\lambda}$.
\hbn
In order to show integrality of 
$n^{(\lambda)\;J}_{\sfata_\pi\sfatb_\sigma}$, 
let us first assume that $C^{\pi*\sigma}_\lambda$ coincides with a
single cycle (the $\lambda^{\rm th}$, say) of $\pi$, $\sigma$ and 
$\pi^{-1}\sigma$, and that $\Lambda_\lambda^* = 2M+1$ is odd. (E.g., 
$\pi = (1\,2\,3),\ \sigma = \pi^{-1}$.) We introduce the objects 
$$
N_{a b}^{(M)\;c} := \sum_i \ {S_{a\, i}\,
S_{b\, i}\,S_{c\, i} \over \bigl(S_{0\,i}\bigr)^{2M+1}}
\eq\affgendef$$
such that, for our simplified situation, $n^{(\lambda)\;J}_{\sfata_\pi
\sfatb_\sigma} = N_{\alpha_\lambda \beta_\lambda}^{(M)\;j_\lambda}$. 
For $M=0$, Verlinde's formula tells us that the numbers in 
\affgendef\ are the fusion rules. For arbitrary $M\in\Z_+$, they 
can be rewritten\footnote{${}^3$}{I am grateful to Jean-Bernard Zuber 
for pointing this out to me.}  as the numbers of chiral 3-point blocks on a 
Riemann surface of genus $M$, 
$$\eqalign{
N_{a b}^{(M)\;c} &= \sum_{{i_1,\ldots,i_{2M+1} \atop j_1,\ldots,j_{M-1}}}
 N^a_{i_1\,i_2}\,N^b_{i_2\,i_3}\,N^c_{i_3\,i_4}\;
\prod_{l=1}^{M-1}\, N^{j_l}_{i_{l+3}\,i_{l+4}}\, 
                    N^{j_{M-l}}_{i_{M+2+l}\,i_{M+3+l}}
\cr
&= \sum_{j_1,\ldots,j_{M-1}}\ {\rm tr}\,\bigl( N_a N_b N_c\, N_{j_1}^2 \cdots 
   N^2_{j_{M-1}} \bigr)
\cr}$$
where in the first line one identifies $i_{2M+2} = i_1$. 
\hbn
Alternatively, $N_{a b}^{(M)\;c} \in \Z$ follows from the 
(affine graded fusion ring) relations 
$$
N_a^{(M)}\; N_b^{(L)} = \sum_c\ N_{a b}^{(K)\;c}\
   N_c^{(M+L-K)} \ \;,
\eq\gradaffring$$
which hold for all $0\leq K \leq M+L$; we have used the matrix notation 
familiar from the fusion rules. Eq.\ \gradaffring\ is easily 
derived from unitarity of the $S$-matrix and the representation 
property \qdimfus. 
This shows that the $N_a^{(M)}$ span a commutative associative ring, which 
moreover is generated by the fusion rules $N_a^{(0)}$ and a single 
additional matrix $N_0^{(1)}$ with integer entries 
$$
N_{0\,a}^{(1)\;b} = \sum_i\; {S_{a\, i}\,S_{b\, i} \over 
    \bigl(S_{0\,i}\bigr)^{2}} = \sum_k\ N_{ab}^k\;{\rm tr}\,N_k\ \ .
$$
\sn
Now let $\pi$ and $\sigma$ be ``in general position''. Then 
$C^{\pi*\sigma}_\lambda$ covers $K_\lambda\!\!\uupi\ \,$\ cycles of $\pi$, 
$K_\lambda\!\!\uusi\,\ $\ cycles of $\sigma$ and 
$K_\lambda\!\!\uupsi\ \,$\ cycles of $\pi^{-1}\sigma$, 
and the numerator of $n^{(\lambda)\;J}_{\sfata_\pi\sfatb_\sigma}$ 
contains a product of $K_\lambda:= K_\lambda\!\!\uupi\ + 
K_\lambda\!\!\uusi\ + K_\lambda\!\!\uupsi\;$\ matrix elements 
of $S$. But as long as 
$$
D_\lambda(\pi,\sigma) := 
\Lambda_\lambda^* - K_\lambda +3\ \in\ 2\Z_+ + 1 \ \ ,
\eq\NKconstr$$
we can apply the relation \qdimfus\ repeatedly ($K_\lambda -3$ times) 
to reduce to the situation above, and the coefficients 
$n^{(\lambda)\;J}_{\sfata_\pi\sfatb_\sigma}$ are products of 
fusion matrices with some $N^{{\scriptscriptstyle (M)}}_a$, 
showing that Cardy's constraints are satisfied for arbitrary 
$\pi$ and $\sigma$. 
\hbn
That the relation \NKconstr\ always holds can be proved by induction 
in $\Lambda_\lambda^*$, starting from $\Lambda_\lambda^* =2$, where 
it is easy to check all possible cases. 
Now assume that \NKconstr\ is true for all permutations $\pi,\;\sigma$ 
such that the $C_\lambda^{\pi*\sigma}$ have length (at most) $\Lambda^* -1$. 
Obviously, it is sufficient to focus on situations where 
$\pi*\sigma$ has just a single orbit $C_{1}^{\pi*\sigma}$
of that maximal length, so we can assume that $\pi,\sigma \in 
S_{\Lambda^*-1}$. In order to increase the orbit length by one, we have to 
pass to $\tilde\pi,\tilde\sigma \in S_{\Lambda^*}$. But with the help of 
transpositions $\tau_{i,j}\,$, every such permutation can be written as 
$$
\tilde\pi = \hat\pi \circ \tau_{i^\pi\!,\Lambda^*}\quad\ \ 
{\rm for\ some}\ \ \pi\in S_{\Lambda^*-1}\ \ {\rm and\ some}\ \ 
i^{\pi} \in \{1,\ldots,\Lambda^*\}
$$
where $\hat\pi$ denotes the trivial extension of $\pi$ to 
$\{1,\dots,\Lambda^*\}$, i.e.\ $\hat\pi(j)=\pi(j)$ for 
$1\leq j\leq \Lambda^*-1$ and $\hat\pi(\Lambda^*)=\Lambda^*$. 
Analogously, we can write $\tilde\sigma= \hat\sigma\circ 
\tau_{i^\sigma\!,\Lambda^{\!{}^*}}$. Except for the trivial case 
$i^\pi=i^\sigma = \Lambda^*$, we have $C^{\tilde\pi*\tilde\sigma}_{1}
= \Lambda^*$. If $i^\pi \neq \Lambda^*$ and $i^\pi\in C_\nu^\pi$, then 
the $\nu^{\rm th}$ cycle of $\tilde\pi$ is obtained from the $\nu^{\rm th}$ 
cycle of $\pi$ by placing $\Lambda^*$ right behind $i^\pi$, and 
$K^{\tilde\pi}_1 = K^\pi_1$. If $i^\pi=\Lambda^*$, then 
$K^{\tilde\pi}_{1} = K^\pi_1 + 1$, but 
$K^{\tilde\sigma}_1 = K^\sigma_1$ since $i^\sigma\neq \Lambda^*$. 
The permutation $\pi^{-1}\sigma$ changes to 
$\tilde\pi^{-1}\tilde\sigma = \tau_{i^\pi\!,\Lambda^{\!{}^*}}\circ 
\pi^{-1}\sigma\circ \tau_{i^\sigma\!,\Lambda^{\!{}^*}}.$ 
If $i^\sigma \neq \Lambda^*$, then $\Lambda^*$ is inserted in an 
existing $(\pi^{-1}\sigma)$-cycle directly behind $i^\sigma$. 
The effect of $\tau_{i^\pi\!,\Lambda^{\!{}^*}}$, assuming 
$i^\pi\neq \Lambda^*$, depends on whether $i^\pi$ is an element 
of that $(\pi^{-1}\sigma)$-cycle or not: In the first case, the cycle 
is split into two (between $i^\pi$ and its predecessor in the 
cycle); in the second case, the $(\pi^{-1}\sigma)$-cycle containing 
$i^\pi$ is joined with the one containing $i^\sigma$. 
Counting the number of $\tilde\pi$-, $\tilde\sigma$- and 
$(\tilde\pi^{-1}\tilde\sigma)$-cycles covered by
$C^{\tilde\pi*\tilde\sigma}_1$, it is easy to see that 
$$
D_1(\tilde\pi,\tilde\sigma) = D_1(\pi,\sigma)\ \; {\rm or}\ \; 
D_1(\pi,\sigma) + 2
$$
for all possible $i^\pi,i^\sigma$ and cycle structures -- so 
$D_1(\tilde\pi,\tilde\sigma)$ is odd as required.  All 
in all, we have shown that \NKconstr\ indeed holds and therefore 
that all pairs of boundary states defined by \permIshi\ and 
\ansatz\ satisfy Cardy's conditions. 
\bn\mn
{\bf Appendix B} 
\mn
We express some of the intersection forms for permutation branes 
on the quintic in terms of charge symmetry generators. 
The fields and states of bulk Gepner models transform under a 
discrete charge symmetry group which, up to $\Z_2$-factors we will 
ignore, is given by $(\Z_5)^4$ in the case of the quintic. The 
generators act on Ishibashi states (for both A- and B-type gluing 
conditions) as follows:
$$
G_j \; |{\fatl} ,{\fatm}\rangle\!\rangle = 
e^{{2\pi i\over 5}m_j}\;|{\fatl} ,{\fatm}\rangle\!\rangle\
$$
for $j=1,\ldots,5$. Since the $\beta$-constraints imply that 
the total charge of each state is integer, one has the 
relation $G_1\,G_2\,G_3\,G_4\,G_5 = 1$. On boundary states, 
the $\Z_5$-generators act by shifting the $M$-labels; we denote 
the generator acting on the label $M_\xi$ by $g_\xi$, where 
the index $\xi$ may stand for the number $\nu$ of a $\pi$-cycle, 
for $[\nu]$ in the B-type cases, etc.
For $\pi$-permuted A-type branes, we have 
$$\eqalign{
&g_\nu\, :\; M_\nu \longmapsto M_\nu + 2 
         \quad\quad {\rm for}\ \;P^\pi > 1\ \ ,
\cr
&\vphantom{\sum}g\, :\; M \longmapsto M + 2\ ,\quad\  
 g'\, :\; M' \longmapsto M' + 1 \quad\quad {\rm for}\ \;P^\pi = 1\ \ .
\cr}$$
The two labels $M$ and $M'$, and likewise the generators $g$ and $g'$, 
are completely independent for $\pi = (1\,2\,3\,4\,5)$, while for 
$P^\pi > 1$ the $\beta$-constraints enforce the relation
$$
g_1^{\Lambda\upi_1}\, g_2^{\Lambda\upi_2}\, \cdots 
       \, g_{P\upi}^{\Lambda\upi_{P\upi}} = 1\ \;. 
\eq\Grelation$$
For B-type boundary states, the labels $M$ and $M_{[\nu]}$ are 
again independent, and so are $g$ and $g_{[\nu]}$, acting as 
$$
g\, :\; M \longmapsto M + 2\ ,\quad\quad\    
 g_{[\nu]}\, :\; M_{[\nu]} \longmapsto M_{[\nu]} + 2 \ \ .
$$
Since the intersection forms $I_{\sfata_\pi\,\tilde{\sfata}_\pi}$ depend 
only on differences of the $M_\xi$- and $\tilde M_\xi$-variables, they 
can be written in terms of $(\Z_5)^R$-generators with 
$$\eqalign{
&\hbox{\rm A-type:}\quad\quad R = P^\pi -1\quad \quad{\rm for}\quad  
P\uupi\; >1
\cr
&\phantom{\hbox{\rm A-type:}}\quad\quad R = 2  \quad\quad \quad \ 
\quad{\rm for}\quad  P\uupi\; = 1
\cr
&\hbox{\rm B-type:}\quad\quad R = 1 + P^\pi_{\rm ev} 
\cr}
$$
with $P^\pi_{\rm ev}$ being the number of even length cycles 
of the permutation $\pi$. The charge symmetry is therefore 
different from the $\pi = {\rm id}$ case analysed in \q{\BDLR}. 
\sn
For $\pi={\rm id}$ branes, closed formulae for intersection forms 
in terms of symmetry generators are easy to write down once the 
SU(2) fusion rules have been expressed through the $g_\nu$. In 
our more general case, analogous expressions have to be found for 
various products of fusion matrices -- a task that becomes more 
and more tedious as the cycle lengths increase. We therefore give 
formulae for A-type intersection forms 
$I^A_{\sfata_\pi\,\tilde{\sfata}_\pi}$ in terms of 
$\Z_5$-generators only for permutations with cycles of length 
up to $\Lambda_\nu = 4$. For  $L_\nu = \tilde L_\nu =0$, the 
intersection forms are 
\def\nsv{\noalign{\vskip.1cm}}
$$\eqalign{
&\pi= {\rm id}\;{\rm :}
\cr
&\phantom{xxxx}I\sim (1-g_1)\,(1-g_2)\,(1-g_3)\,(1-g_4)\,(1-g_5)
\cr\nsv
&\pi = (1)(2)(3)(4\,5)\;{\rm :}
\cr
&\phantom{xxxx}I\sim (1-g_1)\,(1-g_2)\,(1-g_3)\,(1-g_4)(1-3g_4-2g_4^2-g_4^3\,)
\cr\nsv
&\pi = (1)(2\,3)(4\,5)\;{\rm :}
\cr
&\phantom{xxxx}I\sim (1-g_1)\,
    (1-g_2)(1-3g_2-2g_2^2-g_2^3\,)\,(1-g_3)(1-3g_3-2g_3^2-g_3^3\,)
\cr\nsv
&\pi = (1)(2)(3\,4\,5)\;{\rm :}
\cr
&\phantom{xxxx}I\sim (1-g_1)\,(1-g_2)\,(1-g_3)(1+3g_3+g_3^2\,)
\cr\nsv 
&\pi = (1)(2\,3\,4\,5)\;{\rm :}
\cr
&\phantom{xxxx}I\sim (1-g_1)\,(1-g_2^2\,)^2
\cr\nsv
&\pi = (1\,2)(3\,4\,5)\;{\rm :}
\cr
&\phantom{xxxx}I\sim (1-g_1)(1-3g_1-2g_1^2-g_1^3\,)\,
  (1-g_2)(1+3g_2+g_2^2\,)
\cr}$$
In each line, one can eliminate one of the group generators 
by means of \Grelation. Computing $I_{\sfata_\pi\,\tilde{\sfata}_\pi}$
for higher $L$-values, one finds the same behaviour as  in \q{\BDLR}:\ 
for each $L_\nu$ or $\tilde L_\nu$ that is raised from 0 to 1, the 
$g_\nu$-dependent factor has to be multiplied by 
$(g_\nu^\oh + g_\nu^{-\oh}\,)$. 
\sn
For B-type permutation branes, the intersection forms 
$I^B_{\sfata_\pi\,\tilde{\sfata}_\pi}$  
have the following form -- up to cycle length $\Lambda_\nu = 3$, and 
again with $L_\nu = \tilde L_\nu =0$:
\vskip.2cm
\halign{\qquad\qquad $\, # $\qquad \ \hfil & $ # $ \hfill  \cr
\pi= {\rm id}\;{\rm :}
&I\sim\ (1-g)^5 
\cr\nsv\nsv
\pi = (1)(2)(3)(4\,5)\;{\rm :}
&I\sim\ g (1-g)^3 \cdot N(g_{[4]})
\cr\nsv\nsv
\pi = (1)(2\,3)(4\,5)\;{\rm :}
&I\sim\  g^2 (1-g)  \cdot N(g_{[2]}) \cdot N(g_{[3]})\,
\cr\nsv\nsv
\pi = (1)(2)(3\,4\,5)\;{\rm :}
&I\sim\  g^2 (1-g) 
\cr\nsv\nsv 
\pi = (1\,2)(3\,4\,5)\;{\rm :}
&I\sim\  g (1-g)(1+3g+g^2\,)  \cdot N(g_{[1]})
\cr}
\mn
We have used the abbreviation $N(g_{[\nu]}) :=
1 + g_{[\nu]} + g_{[\nu]}^2  + g_{[\nu]}^3 + g_{[\nu]}^4$ 
for the $g_{[\nu]}$-dependent contribution to 
the intersection forms. The form of $N(g_{[\nu]})$ simply 
means that the intersection of two boundary states 
depends only on their $M$-labels and not on the labels $M_{[\nu]}$
associated with cycles of even length; cf.\ to the analogous observation 
made in \q{\BDi} in the context of torsion branes. 
\sn
In order to obtain intersection forms for boundary states where an 
$L_\nu$ or $\tilde L_\nu$ is raised from 0 to 1, one multiplies the 
$g$-dependent part by $(g^\oh + g^{-\oh})$ and, if $\Lambda_\nu$ 
is even, $N(g_{[\nu]})$ by $(g_{[\nu]}^\oh + g_{[\nu]}^{-\oh})$.
\mn
One observes that $g$-dependent factors in these B-type intersection 
forms $I^B_\pi$ (with all $L$-labels zero) can be obtained from the 
A-type intersection forms $I_\pi^A$ for the same permutation $\pi$ 
by replacing all the $g_i$ 
by the single $\Z_5$-generator $g$, up to an overall multiplicity 
which we left undetermined anyway. Thus, the natural conjecture 
for $I^B_{\sfata_\pi\,\tilde{\sfata}_\pi}$  with $\pi = (1)(2\,3\,4\,5)$ is 
$$
I \sim\ g(1+2g-2g^2-g^3\,)
$$ 
up to a (probably again irrelevant) factor depending on $g_{[2]}$. 
This could of course be checked directly, by starting from \intpipiB\ 
and going through some rather tedious combinatorics. 
\mn
One can now in principle follow the methods of \q{\BDLR} and compare 
the $g$-parts of $I^\pi_B$ to the geometric B-type intersection form 
$I^{\rm geo}_B$ of even-dimensional cycles, which at the Gepner point 
is given by 
$$
I^{\rm geo}_B = -g\,(1-g)^3\ \ ,
$$
see \q{\BDLR}. Up to overall normalisation, we 
find that $I^\pi_B \sim m_\pi(g)\; I^{\rm geo}_B\; m_\pi(g^{-1})$
with $m_\pi(g) = 1-g$ for $\pi = {\rm id}$ as in \q{\BDLR} and 
$m_\pi(g) = 1$ for $\pi = (1)\,(2)\,(3)\,(4\,5)$. 
\bn\bn
{\bf References}
\sn
\noindent\def\bf{\bfk}\def\sl{\itk}
{\baselineskip=10pt {\klein 
\def\q#1{\vskip2.5pt\noindent\item{{\lb\klein #1\rb}}}
\q{\Caone} J.L.\ Cardy, {\sl Conformal invariance and surface
   critical behavior}, Nucl.\ Phys.\ B{\bf240} (1984) 514; \quad
  {\sl Effect of boundary conditions on the
 operator content of two-dimensional conformally invariant theories},
 Nucl.\ Phys.\ B{\bf275} (1986) 200  
\q{\Ca} J.L.\ Cardy, {\sl Boundary conditions, fusion rules
 and the Verlinde formula}, Nucl.\ Phys.\ B{\bf324} (1989) 581 
\q{\Lew}  D.C.\ Lewellen, {\sl Sewing constraints for conformal 
 field theories on surfaces with boundaries}, Nucl.\ Phys.\ B{\bf372} 
  (1992) 654
\q{\CaL}  J.L.\ Cardy, D.C.\ Lewellen, {\sl Bulk and boundary operators
 in conformal field theory}, Phys.\ Lett.\ B{\bf259} (1991) 274   
\q{\PSS}  G.\ Pradisi, A.\ Sagnotti, Y.S.\ Stanev, {\sl Completeness 
   conditions for boundary operators in 2d conformal field theory}, 
   Phys.\ Lett.\ B{\bf381} (1996) 97, hep-th/9603097
\q{\Run} I.\ Runkel, {\sl Boundary structure constants for the A-series 
  Virasoro minimal models}, Nucl.\ Phys.\ B{\bf549} (1999) 563, 
  hep-th/9811178; \quad {\sl Structure constants for the D-series 
  Virasoro minimal models}, Nucl.\ Phys.\ B{\bf579} (1999) 561, 
  hep-th/9908046
\q{\BPPZ} R.E.\ Behrend, P.A.\ Pearce, V.B.\ Petkova, J.-B.\ Zuber,  
  {\sl Boundary conditions in rational conformal field theories}, 
  Nucl.\ Phys.\ B{\bf570} (2000) 525 and B{\bf579} (2000) 707, 
  hep-th/9908036 
\q{\FFFS}G.\ Felder, J.\ Fr\"ohlich, J.\ Fuchs, C.\ Schweigert, 
  {\sl Conformal boundary conditions and three-dimensional topological 
  field theory},  Phys.\ Rev.\ Lett.\ {\bf84} (2000) 1659, 
  hep-th/9909140;\ \  {\sl Correlation functions and 
  boundary conditions in RCFT and three-dimensional topology}, 
  hep-th/9912239
\q{\JoseSonia} J.\ Figueroa-O'Farrill, S.\ Stanciu, {\sl D-branes in}
 ${\scriptstyle AdS_3\times S^3 \times S^3 \times S^1}$, 
 J.\ High Energy Phys.\ 0004 (2000) 005  
\q{\Costasetal} C.\ Bachas, J.\ de Boer, R.\ Dijkgraaf, H.\ Ooguri, 
 {\sl Permeable conformal walls and holography}, J.\ High Energy Phys.\ 
 0206 (2002) 027,  hep-th/0111210
\q{\FS} J.\ Fuchs, C.\ Schweigert, {\sl A classifying algebra for 
  boundary conditions}, Phys.\ Lett.\ B{\bf414} (1997) 251, hep-th/9708141
\q{\RSm}A.\ Recknagel, V.\ Schomerus, {\sl Boundary deformation 
   theory and moduli spaces of D-branes}, Nucl.\ Phys.\ B{\bf 545} (1999) 
   233, hep-th/9811237
\q{\KS} A.\ Klemm, M.G.\ Schmidt, {\sl Orbifolds by cyclic 
  permutations of tensor product conformal field theories}, 
  Phys.\ Lett.\ B{\bf245} (1990) 53 
\q{\BHS} L.~Borisov, M.B.~Halpern, C.~Schweigert, {\sl Systematic 
  approach to cyclic orbifolds}, Int.\ J.\ Mod.\ Phys.\ A{\bf13} 
  (1998) 125, hep-th/9701061
\q{\MattTerr} M.R.\ Gaberdiel, T.\ Gannon, {\sl Boundary states for WZW 
  models}, hep-th/0202067
\q{\Ban} P.~Bantay, {\sl Orbifolds, Hopf algebras and the moonshine},
  Lett.\ Math.\ Phys.\ {\bf22} (1991) 187; \quad {\sl Orbifolds and 
  Hopf Algebras}, Phys.\ Lett.\ B{\bf245} (1990) 477;\quad 
  {\sl Characters and modular properties of permutation orbifolds},
  Phys.\ Lett.\ B{\bf419} (1998) 175, hep-th/9708120;\quad
  {\sl Permutation orbifolds}, Nucl.\ Phys.\ B{\bf633} (2002) 365, 
  hep-th/9910079
\q{\FM}H.\ Fuji, Y.\ Matsuo, {\sl Open string on symmetric product}, 
  Int.\ J.\ Mod.\ Phys.\ A{\bf16} (2001) 557, hep-th/0005111
\q{\ncg} V.\ Schomerus,  {\sl D-branes and deformation quantization}, 
    J.\ High Energy Phys.\  06 (1999) 030, hep-th/9903205
\noindent\item{} A.Yu.\ Alekseev, A.\ Recknagel, V.\ Schomerus, 
  {\sl Non-commutative world-volume geometries: branes on SU(2) and 
     fuzzy spheres}, J.\ High Energy Phys.\ 09 (1999) 023, hep-th/9908040;
  \quad  {\sl Brane dynamics in background fluxes and non-commutative 
   geometry}, J.\ High Energy Phys.\ 0005 (2000) 010, hep-th/0003187
\noindent\item{}  N.\ Seiberg, E.\ Witten, {\sl String theory and 
  noncommutative geometry}, J.\ High Energy Phys.\ 09 (1999) 032, 
  hep-th/9908142
\q{\Antonetal} A.Yu.\ Alekseev, S.\ Fredenhagen, T.\ Quella, V.\ Schomerus, 
  {\sl  Non-commutative gauge theory of twisted D-branes}, hep-th/0205123
\q{\Gep}  D.\ Gepner, {\sl Space-time supersymmetry in compactified
 string theory and superconformal models}, Nucl.\ Phys.\ B{\bf296}
 (1987) 757
\q{\FKS}
  J.~Fuchs, A.~Klemm, M.G.~Schmidt, {\sl Orbifolds by cyclic 
  permutations in Gepner type superstrings and in the corresponding 
  Calabi-Yau manifolds}, Ann.\ Phys.\ {\bf 214} (1992) 221
\q{\OOY} H.\ Ooguri, Y.\ Oz, Z.\ Yin, {\sl D-branes on
  Calabi-Yau spaces and their mirrors}, Nucl.\ Phys.\ B{\bf477}
  (1996) 407, hep-th/9606112
\q{\RSg} A.\ Recknagel, V.\ Schomerus, {\sl D-branes in Gepner 
  models}, Nucl.\ Phys.\ B{\bf531} (1998) 185, hep-th/9712186
\q{\BDLR} I.\ Brunner, M.R.\ Douglas, A.\ Lawrence, C.\ R\"omelsberger, 
  {\sl D-branes on the quintic}, J.\ High Energy Phys.\ 0008 (2000) 
  015, hep-th/9906200
\q{\CY} D.-E.\ Diaconescu, C.\ R\"omelsberger, {\sl D-branes 
  and bundles on elliptic fibrations}, Nucl.\ Phys.\ B{\bf574} 
  (2000) 245, hep-th/9910172
\noindent\item{} P.\ Kaste, W.\ Lerche, C.A.\ L\"utken and J.\ Walcher,
  {\sl D-branes on K3-fibrations}, Nucl.\ Phys.\ B{\bf582} (2000) 203, 
  hep-th/9912147
\noindent\item{} E.\ Scheidegger, {\sl D-branes on some one- and 
  two-parameter Calabi-Yau hypersurfaces}, 
  J.\ High Energy Phys.\ 0004 (2000) 003, hep-th/9912188
\noindent\item{} M.\ Naka, M.\ Nozaki, {\sl Boundary states in Gepner 
 models}, J.\ High Energy Phys.\ 0005 (2000) 027, hep-th/0001037 
\q{\DouCat} M.R.\ Douglas,  {\sl D-branes, categories and N=1 
  supersymmetry}, hep-th/0011017
\q{\BS} I.\ Brunner, V.\ Schomerus, {\sl D-branes at singular 
  curves of Calabi-Yau compactifications}, J.\ High Energy Phys.\ 
  0004 (2000) 020, hep-th/0001132
\q{\FSW}J.\ Fuchs, C.\ Schweigert, J.\ Walcher, {\sl Projections in string 
  theory and boundary states for Gepner models}, Nucl.\ Phys.\  
  B{\bf588} (2000) 110, hep-th/0003298 
\q{\BDi} I.~Brunner and J.~Distler, {\sl Torsion D-branes in 
  nongeometrical phases}, hep-th/0102018
\q{\DFi} M.R.\ Douglas, B.\ Fiol,  {\sl D-branes and discrete torsion II}, 
  hep/th/9903031
\q{\MattSaku} M.R.\ Gaberdiel, S.\ Sch\"afer-Nameki, {\sl D-branes in an 
  asymmetric orbifold}, to appear
\q{\PZ} V.\ Petkova, J.-B.\ Zuber, {\sl BCFT:\ from the boundary to the bulk}, 
 Proc.\ TMR meeting Paris 2000, hep-th/0009219; \quad
 {\sl Generalised twisted partition functions}, 
  Phys.\ Lett.\ B{\bf504} (2001) 157, hep-th/0011021; \quad 
  {\sl The many faces of Ocneanu cells},  Nucl.\ Phys.\ B{\bf603} (2001) 449,
  hep-th/0101151; \quad 
  {\sl Conformal boundary conditions and what they teach us}, Proc.\ 
  Budapest 2000, hep-th/0103007
\q{\Scheid} E.\ Scheidegger, {\sl On D0-branes in Gepner models}, hep-th/0109013 
\q{\GJS}S.\ Govindarajan, T.\ Jayaraman, T.\ Sarkar,
  {\sl Worldsheet approaches to D-branes on supersymmetric cycles}, 
  Nucl.\ Phys.\ B{\bf580} (2000) 519, hep-th/9907131
\q{\HIV} K.\ Hori, A.\ Iqbal, C.\ Vafa, {\sl D-Branes And Mirror Symmetry},  
  hep-th/0005247
\q{\Kenn} K.D.\ Kennaway, {\sl A geometrical construction of rational boundary 
   states in linear sigma models}, hep-th/0203266
\smallskip}}
\bye